\documentclass[preprint,5p,times,twocolumn]{elsarticle}
\biboptions{square,sort&compress}
\DeclareUnicodeCharacter{2061}{}

\usepackage{amsmath}

\usepackage{amssymb}
\usepackage{lipsum}
\usepackage{amsthm}
\usepackage{hyperref}
\usepackage{diagbox}

\usepackage[caption=false]{subfig}
\usepackage{floatrow}
\usepackage{dblfloatfix}    
\usepackage{algorithm}
\usepackage{algpseudocode}
\usepackage[switch]{lineno} 

\journal{Physics Letters A}

\begin{document}

\begin{frontmatter}



\title{Optimal Qubit Mapping Search for Encoding Classical Data into Matrix Product State Representation with Minimal Loss}


\author[first]{Hyungjun Jeon}
\author[first]{Kyungmin Lee}
\affiliation[first]{Department of Computer Science and Engineering, Seoul National University, Seoul 08826, Republic of Korea}

\author[second]{Dongkyu Lee}
\author[second]{Bongsang Kim}
\affiliation[second]{Quantum AI Dept, AI Lab, CTO, LG Electronics, Seoul 06772, Republic of Korea}
\author[first,c,d,e,f,g]{Taehyun Kim}
\ead{taehyun@snu.ac.kr}
\affiliation[c]{Department of Computer Science and Engineering, Seoul National University, Seoul 08826, Republic of Korea}
\affiliation[d]{Automation and System Research Institute, Seoul National University, Seoul 08826, Republic of Korea}
\affiliation[e]{Inter-university Semiconductor Research Center, Seoul National University, Seoul 08826, Republic of Korea}
\affiliation[f]{Institute of Computer Technology, Seoul National University, Seoul 08826, Republic of Korea}
\affiliation[g]{Institute of Applied Physics, Seoul National University, Seoul 08826, Republic of Korea}

\begin{abstract}
Matrix product state (MPS) offers a framework for encoding classical data into quantum states, enabling the efficient utilization of quantum resources for data representation and processing. This research paper investigates techniques to enhance the efficiency and accuracy of MPS representations specifically designed for encoding classical data. Based on the observations that MPS truncation error depends on the pattern of the classical data, we devised an algorithm that finds optimal qubit mapping for given classical data, thereby improving the efficiency and fidelity of the MPS representation. Furthermore, we evaluate the impact of the optimized MPS in the context of quantum classifiers, demonstrating their enhanced performance compared to the conventional mapping. This improvement confirms the efficacy of the proposed techniques for encoding classical data into quantum states. MPS representation combined with optimal qubit mapping can pave a new way for more efficient and accurate quantum data representation and processing.
\end{abstract}


\end{frontmatter}




\section{Introduction}
The dimension of the Hilbert space represented by a quantum computer grows exponentially as a function of the number of qubits, which makes quantum computers operate in fundamentally different ways. Various quantum algorithms are suggested that can solve computational tasks exponentially faster than their classical counterpart.  These algorithms include quantum Fourier transform \cite{RN1}, factoring algorithm \cite{RN2}, solving systems of linear equations \cite{RN3, duan2020survey}, quantum support vector machine \cite{RN4}, quantum principal component analysis \cite{RN5}, and quantum convolutional neural network \cite{RN6}. Except so-called Shor’s algorithm which prepares the quantum state using a mathematical relation, those algorithms usually come with an assumption of black boxes for quantum state preparation \cite{RN7}. In most research, it is assumed that these black boxes have polynomial computational complexity with respect to the number of qubits. In some cases, it has been demonstrated that the quantum advantage of algorithms can be attributed to the assumption of state preparation \cite{RN7}. With an access to such kind of black box and algorithms’ underlying assumptions, it was shown that classical computers can perform the same tasks only polynomially worse than the quantum computers \cite{RN8}.

On the other hand, it is well-known that preparing arbitrary n-qubit quantum states requires exponential overhead $O(2^n)$ \cite{RN9,RN10}. For example, Araujo et al. \cite{RN11} and Ghosh \cite{RN12} proposed quantum circuits for state preparation that require exponential depth while Zhang et al. suggested a polynomial-depth circuit but only with exponential number of auxiliary qubits \cite{RN13}. Other approaches assume the use of the quantum random access memory (QRAM) \cite{RN14, RN15, RN16} which generally requires exponential number of memory qubits \cite{RN17, RN18}. Note that all the previously mentioned quantum circuits prepare exact quantum states, and they do not make any assumptions about the data that need to be encoded. However, real-world data often contain correlations or patterns. In classical information processing, redundancies play a crucial role in the efficient encoding and storage of data, as they allow for compression and encoding of the data in a more compact format. For lossless compression, algorithms such as Huffman coding \cite{RN19}, Lempel-Ziv-Welch (LZW) coding \cite{RN20}, and arithmetic coding \cite{RN21} are some of the methods that utilize redundancies to achieve efficient data compression. For lossy compression, techniques such as discrete cosine transform (DCT) \cite{RN22}, wavelet transform \cite{RN23}, vector quantization \cite{RN24}, fractal compression \cite{RN25}, and low-rank approximation \cite{RN26} are commonly used to achieve efficient compression by discarding some of the information in the original data, while maintaining the essential features. These algorithms are used in applications such as image and video compression, where a reduction in the storage or transmission is essential, and some loss of information can be tolerated. Among these techniques, low-rank approximation is a mathematical tool that can be applied to various areas such as numerical linear algebra \cite{RN27}, machine learning \cite{RN28}, signal processing \cite{RN29}, and quantum physics \cite{RN30} where a given matrix is represented as a product of two or more lower rank matrices, with the aim of reducing the computational complexity and storage requirements of the original matrix. By approximating a high-rank matrix with low-rank ones, it is often possible to obtain an accurate representation of the original data with small computational cost and memory requirements \cite{RN28}.

In the context of quantum mechanics, matrix product state (MPS) is a mathematical framework used to represent quantum states with local entanglement characterized by low-rank matrices \cite{RN30}. MPS representation of quantum states requires only polynomial growth in the number of components with respect to the number of qubits \cite{RN31}. This contrasts with the general state vector representation which requires exponential number of components. A parameter called bond dimension determines the capability of MPS in representing quantum states. It regulates the correlation length of MPS as well as its complexity. MPS has also proven to be successful in applications requiring efficient numerical calculation of quantum states such as quantum simulation \cite{RN32, RN33, RN34} and quantum chemistry \cite{RN35}. It was shown that MPS with low bond dimension can be easily mapped to a quantum circuit \cite{RN36, RN37} which allows us to implement MPS efficiently on a quantum computer \cite{RN38}. Previous research also showed that images such as modified National Institute of Standards and Technology database (MNIST) \cite{RN39} and Fashion-MNIST \cite{RN40} can be encoded into MPS with low bond dimension \cite{RN41, RN42}. Additionally, MPS has been utilized as a powerful mathematical framework for tackling various classical machine learning problems beyond quantum machine learning, including but not limited to classification \cite{RN28, RN43} and generative model \cite{RN44}. 

Similar to the classical compression, it is conceivable that a method exists to transform quantum data into a more compressed format for encoding purposes. This would involve finding a way to represent the data using a smaller number of quantum gates, without losing crucial information about the original data. Such a method could be of great value in various applications of quantum information processing, where efficient encoding of data is essential. This study introduces a novel technique aimed at improving the fidelity of the MPS. Specifically, we propose a method that involves permuting the qubits used by encoding circuits. To verify the effectiveness of the our method, we apply this approach to the encoding of data, specifically the MNIST, Fashion-MNIST datasets, pre-processed Canadian Institute for Advanced Research (CIFAR-10) \cite{RN45}, as well as a dense layer of a neural network. Our approach leads to an improvement in accuracy of quantum state encoding, which also turned into enhancement in training two types of quantum classifiers, thereby demonstrating the overall utility of the proposed method.

\section{MPS optimization with permutation}
We explore the optimization algorithm for MPS, and based on its mathematical structure, we introduce a search algorithm for qubit mapping that minimizes MPS truncation errors. Furthermore, we identify symmetries in the permutation of qubits and leverage them to accelerate the search process. We apply our proposed algorithm to encode MPS for various datasets, including the MNIST, Fashion-MNIST, and pre-processed CIFAR-10 datasets. The results show that our algorithm significantly improved the fidelity of encoded quantum states compared to the standard MPS encoding methods \cite{RN46}. Additionally, our algorithm can achieve these improvements without significant additional overhead in quantum circuits, making it a promising technique for various quantum algorithms such as quantum machine learning, quantum optimization, and quantum simulations.

\subsection{Theory}
An MPS represents a quantum state as a product of a sequence of matrices, each of which describes the state of a few adjacent qubits. This representation is particularly useful for states with local entanglement, where the correlations between different parts of the system decay rapidly with distance \cite{RN31}. The MPS provides an efficient way to encode and manipulate quantum states. The bond dimension, denoted as $\chi$, serves as an intrinsic parameter that determines the dimensions of internal indices of the MPS. It dictates the entanglement length of the states that can be effectively represented by the MPS. However, it is worth noting that increasing the bond dimension results in higher computational costs and memory requirements, thus posing practical limitations in terms of computational resources for MPS-based algorithms.

A general {\it n}-qubit quantum state can be parameterized with $2^n$ coefficients as follows:
\begin{equation}
\lvert\psi\rangle = \sum_{s_1,s_2,...,s_n=0}^1 {T^{s_1s_2...s_n}\lvert s_1 \rangle \otimes \lvert s_2 \rangle \otimes ... \otimes \lvert s_n \rangle}, 
\end{equation}
where $T^{s_1 s_2…s_n}$ is a tensor that represent an {\it n}-qubit quantum state with index $s_j$ corresponding to the quantum state of the {\it j}-th qubit.

An MPS representation approximates the coefficient tensor $T^{s_1 s_2…s_n}$ of an {\it n}-qubit quantum state to $\Tilde{T}$ by decomposing it into local tensors with low bond dimensions as follows:

\begin{equation}
    \Tilde{T} = \sum_{\alpha_1,\alpha_2,...,\alpha_n=1}^{\chi} {{A_1}_{\alpha_1}^{s_1} {A_2}_{\alpha_1\alpha_2}^{s_2} ... {A_{n-1}}_{\alpha_{n-1}\alpha_{n}}^{s_{n-1}} {A_n}_{\alpha_n}^{s_n}}
\end{equation}
where $\alpha_i$ is an inner index and $A_i$ is a local tensor.

Generally, the number of parameters increases as $O(n\chi^2)$ and computational complexity of computing expectation values increases as $O(n\chi^3)$ \cite{RN47}.

For a given quantum state, there exists an efficient algorithm that constructs an MPS with a given bond dimension, which has the smallest distance in terms of Frobenius distance \cite{RN46} that can be generalized for a tensor as
\begin{equation}
    \lvert\lvert \Tilde{T} - T\rvert\rvert_F \equiv \sqrt{\sum_{s_1,s_2,...,s_n=0}^1 {\lvert\Tilde{T}^{s_1s_2...s_n}-T^{s_1s_2...s_n}\rvert^2}},
\end{equation}

The algorithm works by iteratively performing singular value decomposition (SVD) and truncate singular values similar to PCA \cite{RN48} until the desired bond dimension is reached.

\begin{center}
\begin{tabular}{p{0.88\linewidth}l}
\hline
Algorithm 1: MPS-SVD \cite{RN46}\\
\hline
{\bf Input:} \\
\hspace{2mm} \textit{n}-qubit quantum state\\
    \hspace{6mm}${\bf a} \xleftarrow{} (a_1,a_2,...,a_{2^n})$\\
\hspace{2mm} bond dimension $\chi$ \\
{\bf Output:} \\
\hspace{2mm} Optimal MPS representation of the quantum state with bond dimension $\chi$ \\
{\bf Initialize:} \\
\hspace{2mm} factors $\xleftarrow{}$ [] \\
1: {\bf while} len({\bf a})  $>$ 2$\chi$ {\bf do} \\
\hspace{2mm} {\bf a} $\xleftarrow{}$ reshape({\bf a}, [2$\chi$, $\frac{len({\bf a})}{2\chi}$]) \\
\hspace{2mm} U, $\Lambda$, $V^\dag$ $\xleftarrow{}$ svd({\bf a}) \\
\hspace{2mm} Truncate U, $\Lambda$, $V^\dag$ up to $\chi$ \\
\hspace{4mm} Append U to \textit{factors} \\
\hspace{4mm} {\bf a} $\xleftarrow{}$ flatten($\Lambda V^\dag$) \\
2: {\bf end}\\
3: {\bf return} factors\\
\hline
\end{tabular}
\end{center}

MPS-SVD algorithm constructs an optimal MPS for a given state. However, the accuracy of the MPS depends heavily on the entanglement structure of the state being encoded \cite{RN46}. It is known that MPS with low bond dimension corresponds to states with local entanglement. Therefore, by permuting the order of the qubits to change the entanglement structure to be more local, we can anticipate a reduction in the truncation error.

\subsection{Finding the optimal permutation}

As previously mentioned, a state exhibiting a local entanglement structure is expected to yield a low truncation error when employing the algorithm outlined above. In this context, the truncation error can be utilized as a cost, and graph search algorithms \cite{RN49, RN50} can be employed to search for the optimal permutation of qubits. By incorporating truncation error as a cost in the search process, our algorithm can be enhanced through the consideration of the approximated state’s quality in the MPS representation.

Since every factor in the MPS decomposition, except for the last one, is unitary, the total truncation error incurred by the algorithm is the sum of the individual truncation errors obtained from each SVD truncation step \cite{RN46}. One notable property of the singular values obtained from the SVD decomposition is their invariance under column exchanges as shown in the following:

\begin{equation}
\begin{split}
&A = U\Lambda V^\dag\\
&AP_{ij} = U\Lambda V^\dag P_{ij} = U'\Lambda V'^\dag    
\end{split}
\end{equation}
where $P_{ij}$ is a column exchange matrix.

This means that changing the order of remaining qubits does not affect the truncation error of the current SVD. As a result, we can compute the truncation error for a particular permutation while considering remaining qubits as a set, regardless of their order.
 
In an MPS representation of the quantum state with multiple qubits, the dimension of the k-th inner index is at most $2^k$. It means that for the first $\lfloor \log_{2} ⁡\chi \rfloor$ qubits and the last $\lfloor \log_{2} ⁡\chi \rfloor$ qubits, the number of singular values is less than or equal to $\chi$. This enables us to treat these qubit arrangements without consideration for their order. This reduction gives ${(\lfloor \log_{2} ⁡\chi \rfloor)!}^2$ speed-up.

Utilizing the aforementioned facts, we can adopt a uniform-cost search algorithm \cite{RN49} to efficiently search for the optimal qubit mapping that lead to the lowest possible truncation error.

\begin{center}
\begin{tabular}{p{0.88\linewidth}l}
\hline
Algorithm 2: MPS-SVD with permutation\\
\hline
{\bf Input:} \\
\hspace{2mm} \textit{n}-qubit quantum state\\
    \hspace{6mm}${\bf a} \xleftarrow{} (a_1,a_2,...,a_{2^n})$\\
\hspace{2mm} number of qubits n\\
\hspace{2mm} bond dimension $\chi$ \\
{\bf Output:} \\
\hspace{2mm} Optimal qubit mapping with the lowest truncation error \\
{\bf Initialize:} \\
\hspace{2mm} min-heap $\xleftarrow{}$ empty heap \\
\hspace{2mm} $x\xleftarrow{} \lfloor{log_2\chi}\rfloor + 1$\\
1: {\bf for} {\it l} {\bf in} {\it combination}({\it n}, {\it x}) {\bf do} \\
\hspace{2mm} \textit{error $\xleftarrow{}$ partial\_truncation\_error(l)} \\
\hspace{2mm} \textit{min-heap.push((error,l))} \\
2: {\bf end}\\
3: {\bf loop do} \\
\hspace{4mm} (error, l) $\xleftarrow{}$ min-heap.pop() \\
\hspace{4mm} \textit{{\bf if} len(l) = n-x {\bf then}} \\
\hspace{6mm} \textit{{\bf return} l} \\
\hspace{4mm} {\bf end} \\
\hspace{4mm} \textit{{\bf for} index {\bf in} not used by l {\bf do}} \\
\hspace{6mm} \textit{l2 $\xleftarrow{}$ copy(l)} \\
\hspace{6mm} \textit{append index to l2} \\
\hspace{6mm} \textit{error $\xleftarrow{}$ error + truncation\_error(l2)}\\
\hspace{6mm} \textit{min-heap.push((error,l2))}\\
\hspace{4mm} {\bf end} \\
4: {\bf end}\\
\hline
\end{tabular}
\end{center}

For a more comprehensive and detailed illustration of the permutation search algorithm employed in our work, refer to \ref{appendixA} where we provide a step-by-step description of the encoding procedure, as well as some additional technical details. The source code of the implementation is available in the github repository \cite{RN51}.

The computational complexity of search for the optimal qubit mapping can become exponential with respect to the number of qubits. In the case of d-dimensional data, the worst-case complexity is 

\begin{equation}
    O((logd)!)=O(d^{loglogd - 1})
\end{equation}

However, with real-world datasets, the effective search space for permutations usually becomes much smaller. Please refer to \ref{appendixA} for numerical results.

\subsection{Image encoding optimization}
 

\begin{figure}
    \centering
    \sidesubfloat[]{\includegraphics[width=0.48\textwidth]{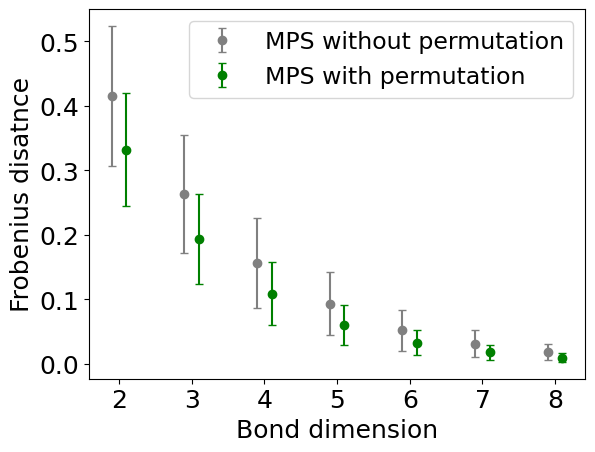}}
    \quad
    \sidesubfloat[]{\includegraphics[width=0.35\textwidth]{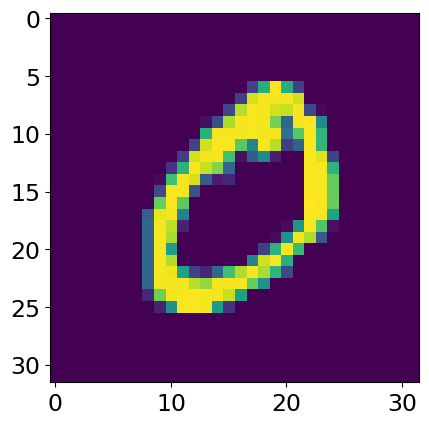}}
    \vskip\baselineskip
    \sidesubfloat[]{\includegraphics[width=0.9\textwidth]{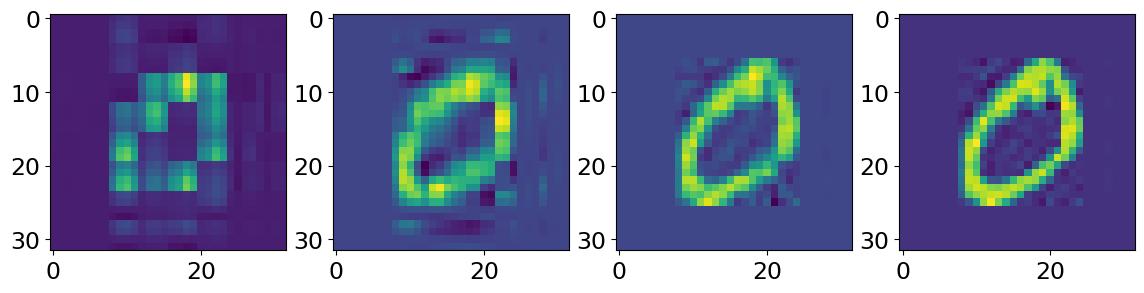}}
    \vskip\baselineskip
    \sidesubfloat[]{\includegraphics[width=0.9\textwidth]{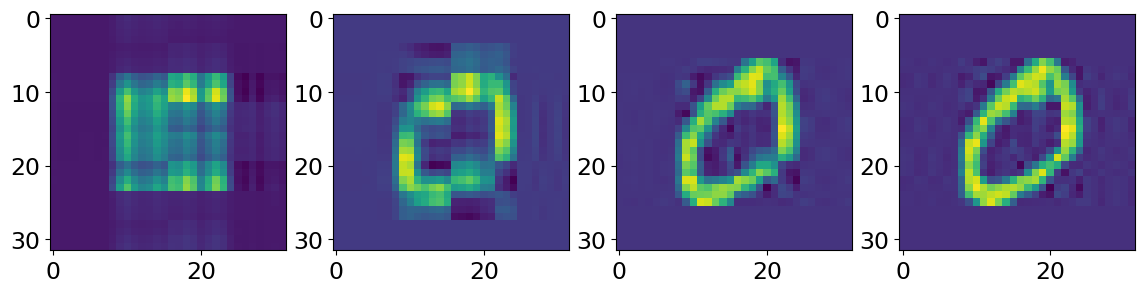}}
    \caption{Comparison of the permuted MPS method and the MPS without permutation for MNIST dataset. (a) Plot of Frobenius distance between the input data and MPS representations with and without permutation as a function of bond dimension. (c), (d) Visual comparison of the MPS states for bond dimensions 2, 4, 6, and 8 with and without permutation corresponding to the original MNIST image data shown in (b).}
    \label{fig:figure1}
\end{figure}

While there are various suggestions for encoding classical data into quantum states \cite{RN52, RN53, li2022quantum}, we use the standard amplitude encoding scheme \cite{RN14} because it is frequently used in quantum image processing \cite{RN54, RN55} due to its efficiency in terms of the number of qubits required to encode data. We first ensure the total number of pixels to be a power of 2 by padding the image if necessary. Then, we flatten the two-dimensional image into a one-dimensional vector using the row-major order, which corresponds to reading the image from left to right and top to bottom. This results in a vector of size $2^n$. We then normalize this vector, which is a standard procedure in quantum machine learning algorithms \cite{RN56}. Finally, this normalized vector is used as an input to the algorithm for MPS with permutation to generate the optimal MPS representation of the image which corresponds to n-qubit quantum states.

To verify the effectiveness of the our algorithm compared to the standard MPS-SVD algorithm \cite{RN46}, we applied both algorithms to three different types of datasets such as MNIST, Fashion-MNIST, and CIFAR-10.

Figure~\ref{fig:figure1} shows the comparison of those two approaches when they are applied to MNIST dataset with {\it n}=10. Figure~\ref{fig:figure1} (a) plots the average Frobenius distances between MPS representations and the original images as a function of bond dimension. Specifically, both data points corresponding to each bond dimension are obtained by applying either method to the individual images in MNIST dataset. For low bond dimensions, the Frobenius distance of the permuted MPS representation is reduced by as much as 32\% compared to the standard MPS schemse, indicating a reduction in truncation error. Note that the difference between two cases gets smaller as we increase the bond dimension because with higher bond dimensions the standard MPS method can also start to capture the underlying remote entanglement structure better, which can be also confirmed by the visual inspection of Fig.~\ref{fig:figure1} (c).

However, the quantum cost to construct an MPS with higher bond dimension increases rapidly, where the MPS with $\chi$ bond dimension requires $(log_2\chi + 1)$-qubit gates \cite{RN57}, eventually reaching the original cost of the exact amplitude encoding, therefore there exists a practical limit to the bond dimension that can be used in practice. 

\begin{figure}[b!]
    \centering
    \sidesubfloat[]{\includegraphics[width=0.48\textwidth]{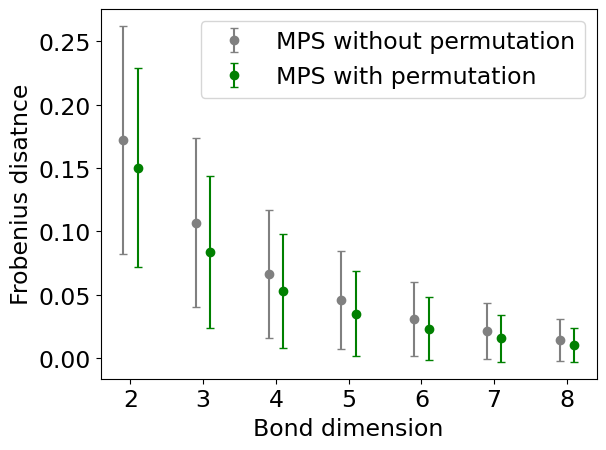}}
    \quad
    \sidesubfloat[]{\includegraphics[width=0.35\textwidth]{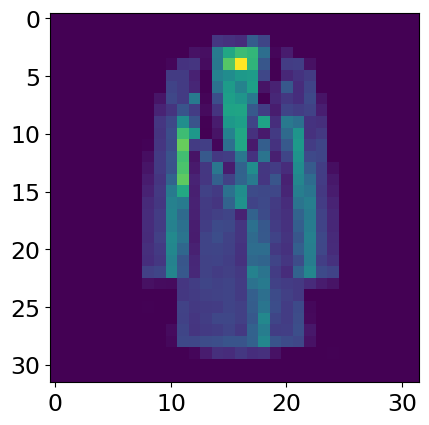}}
    \vskip\baselineskip
    \sidesubfloat[]{\includegraphics[width=0.9\textwidth]{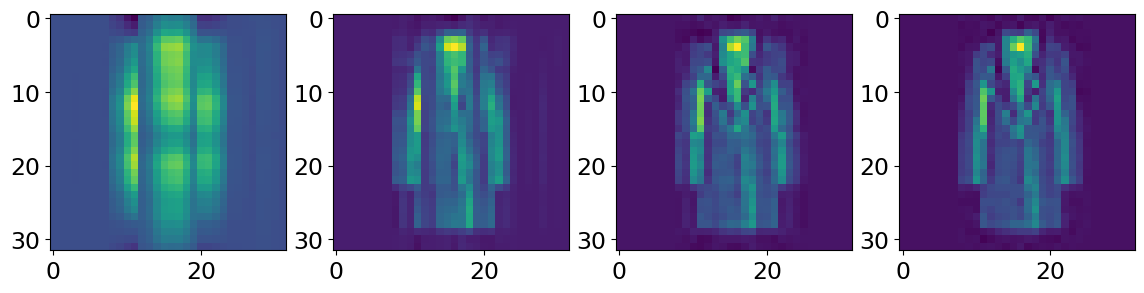}}
    \vskip\baselineskip
    \sidesubfloat[]{\includegraphics[width=0.9\textwidth]{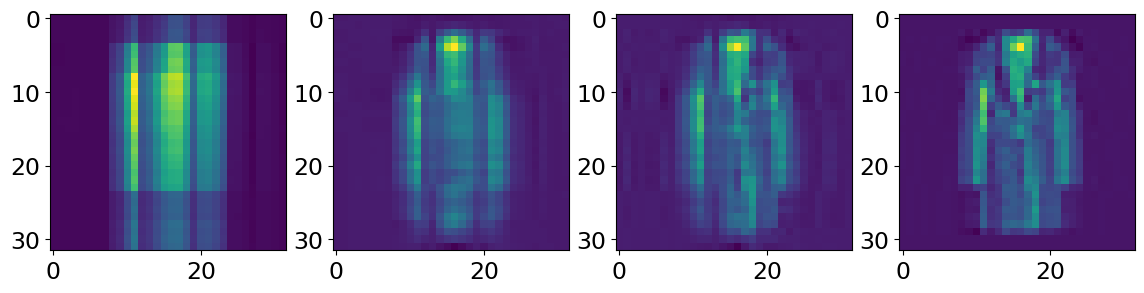}}
    \caption{Comparison of the permuted MPS method and the MPS without permutation for Fashion-MNIST dataset. (a) Plot of Frobenius distance between the input data and MPS representation with and without permutation. (c), (d) Visual comparison of the MPS states for bond dimensions 2, 4, 6, and 8 with and without permutation corresponding to the original Fashion-MNIST image data shown in (b).}
    \label{fig:figure2}
\end{figure}

Figure \ref{fig:figure2} shows the same type of comparison between two MPS representations with Fashion MNIST dataset where similar performance improvement is observed.

As the results indicate, our proposed algorithm has led to a consistent improvement in both the numerical value of the Frobenius distance and the visual appearance of the output images especially for the low bond dimension. Even though the standard deviations shown in both plots might seem larger than the improvements achieved by our algorithm, this can be largely attributed to the relatively broad distribution of error rates among different classes (for more details, refer to \ref{appendixB}). Additionally, the output images appear less distorted, suggesting that our algorithm has successfully captured more relevant features of the input images. Overall, these results demonstrate the effectiveness of our approach for encoding image data into a quantum state using an MPS representation optimized with the qubit permutation search algorithm.

\begin{figure}[t!]
  \centering
  \includegraphics[width=\textwidth]{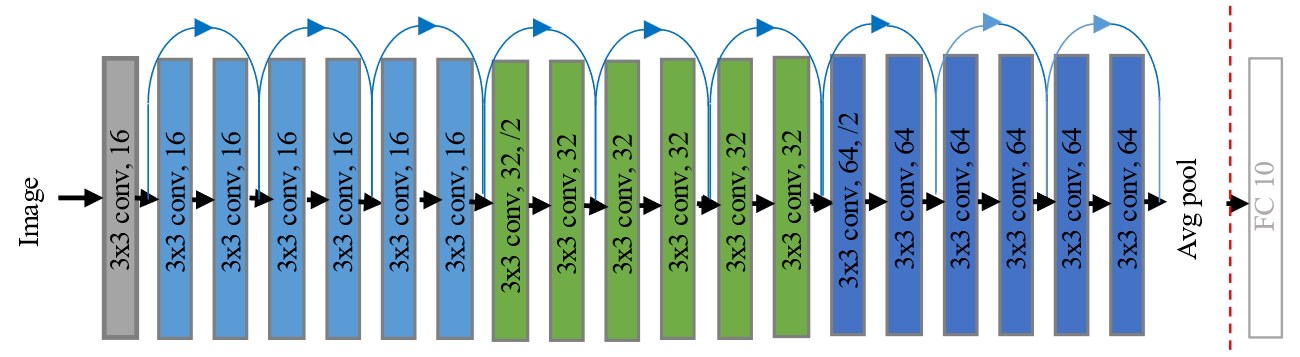}
  \caption{Network architecture used for pre-processing of CIFAR-10 dataset. The circular line denotes the presence of skip connections, while "{\bf n}×{\bf n} conv" indicates the size of the convolution filters, with the number after the comma representing the number of channels. Compared to the standard resnet20 architecture, the fully-connected layer (FC 10) is omitted and instead the input to the FC 10 layer is provided as an input to two MPS algorithms we are comparing.}
  \label{fig:figure3}
\end{figure}

\begin{figure}[b!]
  \centering
  \includegraphics[width=0.8\textwidth]{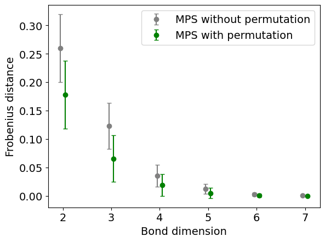}
  \caption{Comparison of Frobenius distance between the input state and MPS representations with and without permutation for pre-processed 64-dimensional CIFAR-10 dataset.}
  \label{fig:figure4}
\end{figure}

In previous studies \cite{RN58, RN59}, it has been common practice to use classical neural networks as a pre-processing layer for quantum neural networks, which is due to their ability to effectively extract features from raw data. These features can then be used as an input for a quantum circuit. By doing so, the quantum neural network can focus on the high-level abstraction of the extracted features, rather than dealing with the raw data directly. To assess the effectiveness of our proposed encoding algorithm even under such kind of scenario, we conduct experiments on the CIFAR-10 dataset using a pre-trained ResNet model \cite{RN60}. We adopted the publicly available simple pre-trained model (resnet20) which achieved the classification accuracy of 92.6\% and can be found at the GitHub repository \cite{RN61} whose network architecture is illustrated in Fig. \ref{fig:figure3}. We use the output of the last pooling layer as the input for quantum encoding. 
  
Figure \ref{fig:figure4} demonstrates that our proposed algorithm achieved a significant reduction in error rate. Furthermore, the error rates exhibit a similar pattern to those observed with previous image datasets, where the improvement is more pronounced in the low bond dimension region. The standard deviations observed in the error rates of the pre-processed datasets are relatively low compared to those of the raw image datasets, which can be attributed to dimensionality reduction. 

Note that the dimension of the output vector from the pre-trained NN is 64, which can be exactly represented by MPS with bond dimension 8. Therefore, we limited our analysis only up to bond dimension 7.

\subsection{Tensorizing pre-trained neural network}
Previous research \cite{RN28} has shown that training a tensor-train in place of a dense layer in a classical neural network can be an efficient approach. By noting that tensor-train has the same mathematical structure as MPS, we can anticipate that a pre-trained dense layer can be approximated more effectively by MPS with permutation.

To benchmark the accuracy of both MPS algorithms, we first train a neural network with a 1024×16 input layer and a 16×10 output layer, and then tensorize the weight matrix of the neural network into an MPS representation, where bond dimensions ranging from 16 to 30 are used to approximate the 1024×16 matrix with MPS using both algorithms. We trained our neural network using the Fashion-MNIST dataset with a batch size of 128, 100 epochs, and the Adam optimizer \cite{RN62} with a learning rate of $10^{-3}$.

\begin{figure}[t]
  \centering
  \includegraphics[width=0.8\textwidth]{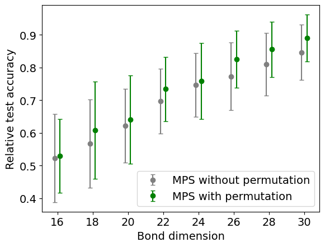}
  \caption{The relative test accuracies of the two MPS layers as a function of bond dimension. The first dense layer is replaced with MPS layers, and the obtained test accuracies are normalized by the test accuracy of the original network.}
  \label{fig:figure5}
\end{figure}

In Fig. \ref{fig:figure5}, the relative test accuracy is plotted as a function of bond dimension for both algorithms, which shows the trade-off between bond dimension and test accuracy and helps us determine the optimal bond dimension for our specific application. Relative test accuracy is defined as follows:

\begin{equation}
    ACC_R \equiv \frac{ACC_T}{ACC}
\end{equation}
where ACC is the test accuracy of original network and $ACC_T$ is the test accuracy of network whose dense layer is replaced with MPS.

The comparison between two methods clearly shows that MPS with permutation is more effective than the standard MPS as a replacement of dense layer by achieving higher accuracy in approximating a part of classical neural network.

\section{Benchmark with quantum classifiers}
To assess the effect of input state quality on quantum information processing tasks, we trained two quantum classifiers - a variational quantum circuit (VQC) classifier and a MPS classifier - using output MPS states of both algorithms. Our results showed a significant increase in the test accuracy of both quantum classifiers that were trained using images encoded by our scheme.

\begin{figure}[t!]
  \centering
  \includegraphics[width=0.9\textwidth]{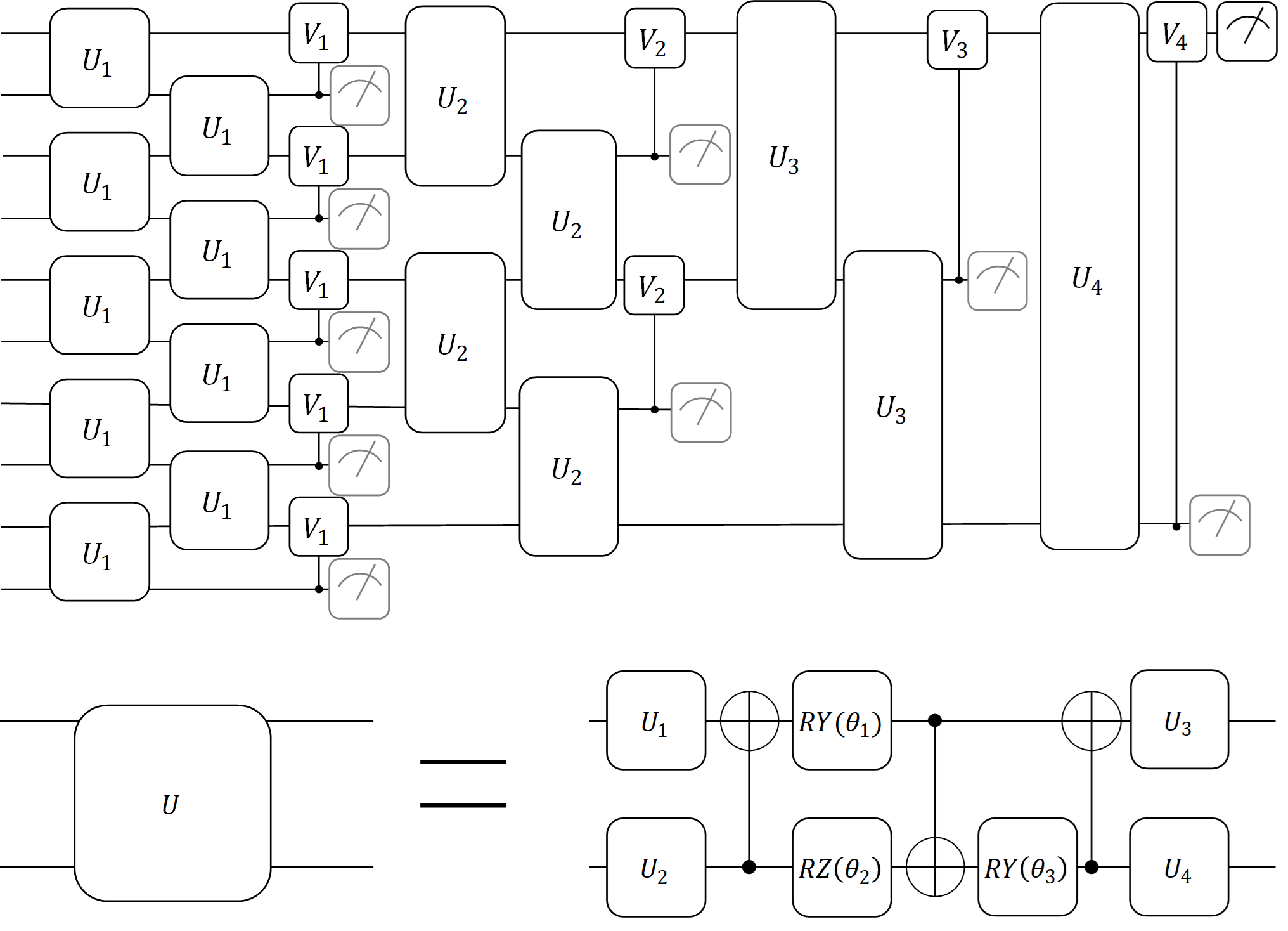}
  \caption{VQC structure employed for the classifier, where the expectation value of the measurement serves as the label for classification.}
  \label{fig:figure6}
\end{figure}

\subsection{VQC classifier}

For benchmarking with a VQC classifier, we attached a quantum convolutional neural network (QCNN) structure \cite{RN6} to the output of the MPS encoding circuit and the expectation value of measurement is used as a probability. The QCNN structure employed for the classifier benchmarking is illustrated at Fig. \ref{fig:figure6}.

In our study, we opted to train our models on a binary classification task using the MNIST dataset \cite{RN63}. To estimate the maximum achievable test accuracy, in addition to both MPS encoding schemes, we included amplitude-encoded states with exact image data in the following comparison. Therefore, there will be three possible encoding schemes for training dataset and also for the test dataset as shown in the Table \ref{tab:table1}. Among all nine combinations, some of the combinations such as the classifier trained by permuted MPS being tested by standard MPS are not considered. 

Our findings indicate that the classifiers trained using our encoding scheme outperformed those trained with the standard MPS independent of the encoding types used to prepare the test input states. For the MPS encoding, we set the bond dimension to 2 and employed the cross-entropy loss function for optimization \cite{RN64}. To carry out the experiments, we utilized the PennyLane software package \cite{RN65} for quantum simulation and trained the entire network utilizing the Adam optimizer with a learning rate of $10^{-2}$. The batch size was set to 128, and we trained the models for a total of 5 epochs.

It is worth mentioning that when testing classifiers trained with MPS images, their accuracy surpasses that of classifiers trained with exact input images, but only if the test set is given in MPS form. This suggests that the classifiers trained with the MPS encoded by both algorithms may have overfitted to the corresponding encoding schemes, which requires further investigation \cite{RN66}. To avoid over-estimation of the power of MPS encoding scheme, we will only consider amplitude-encoded states with exact image data as the test set for the subsequent classifiers. We also evaluated the VQC model's performance by training it with precise images and then testing it using the MPS-encoded test dataset as input. Similarly, the test accuracy exceeded the accuracy obtained with exact test inputs and therefore, we did not include those numbers to avoid potential misunderstanding.

\begin{table}[t!]
\centering
\caption{Test accuracies of VQC classifiers trained by quantum states prepared with different encoding schemes.}
    \small\begin{tabular}
    {p{0.2\linewidth}|p{0.2\linewidth}|p{0.2\linewidth}|p{0.2\linewidth}}
    \diagbox[width=7em]{Train}{Test}  & MPS w/ perm & MPS w/o perm & Exact \\
\hline
     MPS w/ perm    & 95.8±2.3\% & - & 94.6±2.6\% \\
     \hline
     MPS w/o perm &	- & 93.8±9.0\% &	89.5±11.6\% \\
     \hline
     Exact & - & - & 94.6±2.2\% \\
     \hline
    \end{tabular}
    \label{tab:table1}
\end{table}

\subsection{MPS classifier}

\begin{figure*}[!t]
    \centering
    \sidesubfloat[]{\includegraphics[width=0.275\textwidth]{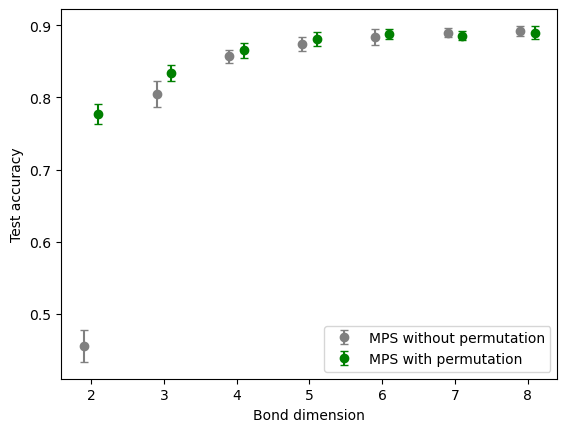}}
    \quad
    \sidesubfloat[]{\includegraphics[width=0.275\textwidth]{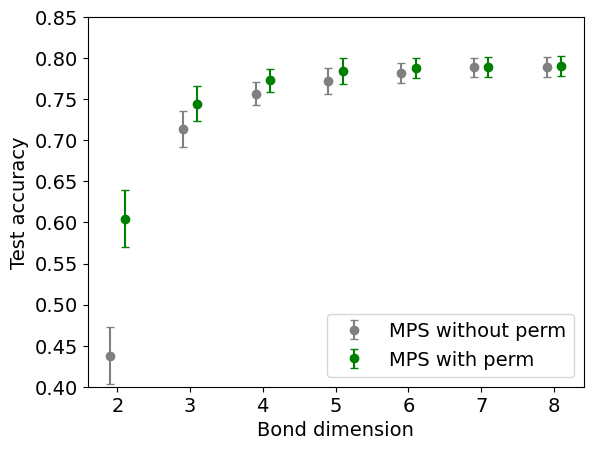}}
    \quad
    \sidesubfloat[]{\includegraphics[width=0.275\textwidth]{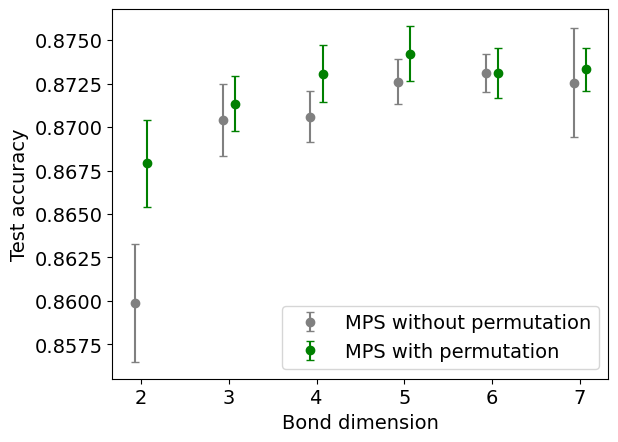}}
    \caption{Test accuracies of MPS classifiers with (a) MNIST, (b) Fashion-MNIST, and (c) pre-processed CIFAR-10 datasets.}
    \label{fig:figure7}
\end{figure*}

Given the higher degree of freedom and easier trainability of MPS classifiers \cite{RN41} compared to the VQC classifiers used in the previous section, we leverage MPS structures to train classifiers for three classification tasks with MNIST, Fashion-MNIST, and pre-processed CIFAR-10 datasets. In our study, we use the MPS classifier with bond dimension 10, and train the MPS classifiers using Adam optimizer with a learning rate of $10^{-2}$. The training is performed with a batch size of 128 and the optimization process runs for 300 epochs.

Figure \ref{fig:figure7} shows that the classifiers trained with our encoding scheme outperform the classifiers trained with the standard MPS. The improvement in performance is the maximum at low bond dimension, which is consistent with the large Frobenius distance gap observed in the previous section II. Additionally, the MPS classifiers appear to be saturated in terms of test accuracy, which we believe is the capacity limitation of the classifiers using the quantum circuits based on MPS structure.

\section{Discussion}
Our study suggests that the proposed encoding scheme improved both the image fidelity in terms of Frobenius distance and the classification accuracy compared to the standard MPS encoding when it is applied to three different types of image datasets. The improvement is pronounced at low bond dimensions, which means that the new scheme will be most useful for the noisy intermediate-scale quantum computer \cite{RN67}.

While the permutation search space might seem to appear to grow exponentially with the number of qubits, we reduced the search space significantly by utilizing the symmetry of the MPS structure under certain permutation, combined with the uniform-cost search. Moreover, our experiments have shown that there exist specific permutations that result in low truncation errors, where only a small portion of the entire permutation space had to be explored while leading to a fast termination.

Even though this study mainly focused on applications of permuted MPS schemes to the quantum machine learning, MPS representation also has a wide range of potential applications such as quantum simulation and quantum chemistry where efficient representation of quantum states is crucial, and MPS combined with efficient search of optimal permutations can lead to significant improvements in many research areas.

\section*{CRediT authorship contribution statement}
\textbf{Hyeongjun Jeon}: Investigation, Visualization, Writing – original draft, Writing – review \& editing
\textbf{Kyungmin Lee}: Writing – review \& editing, Methodology
\textbf{Dongkyu Lee}: Conceptualization
\textbf{Bongsang Kim}: Conceptualization
\textbf{Taehyun Kim}: Writing – review \& editing, Methodology, Supervision

\section*{Declaration of competing interest}
All authors disclosed no relevant relationships.

\section*{Data availability}
Data will be made available on request.

\section*{Acknowledgement}
This work was supported by LG Electronics (Project No. 0418-20220049), the National Research Foundation of Korea (NRF) grant funded by the Korean government (Ministry of Science and ICT, MSIT) (No. 2020R1A2C3005689), and the Institute for Information \& communications Technology Planning \& Evaluation (IITP) grant funded by the Korean government (MSIT) (No. 2022-0-01040).

\appendix
\section{Permutation search}
\label{appendixA}

\begin{figure}[!b]
    \centering
    \sidesubfloat[]{\includegraphics[width=0.4\textwidth]{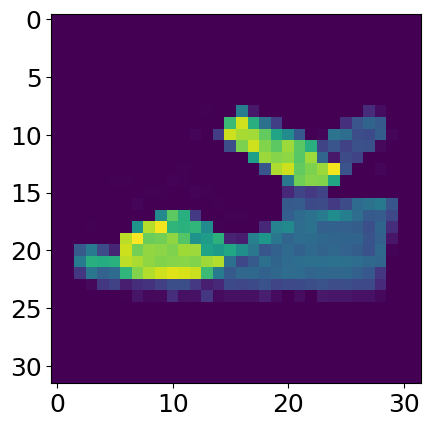}}
    \quad
    \sidesubfloat[]{\includegraphics[width=0.4\textwidth]{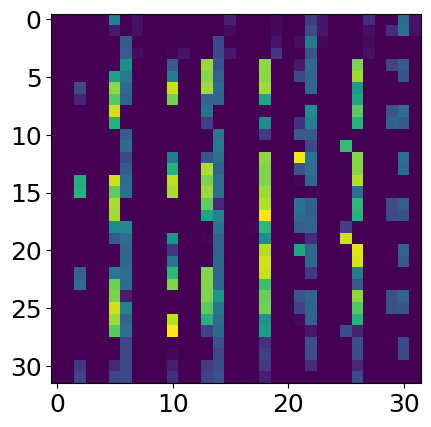}}
    \vskip\baselineskip
    \sidesubfloat[]{\includegraphics[width=0.4\textwidth]{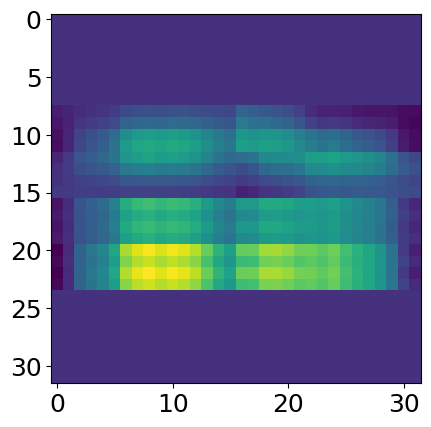}}
    \quad
    \sidesubfloat[]{\includegraphics[width=0.4\textwidth]{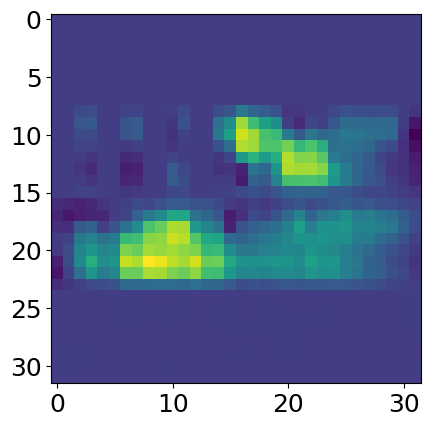}}
    \caption{Visual representation of image data. (a) The original Fashion-MNIST image. (b) Data rearranged by the optimal permutation for MPS encoding. (c) Approximated image of (a) reconstructed by the standard MPS scheme. (d) Approximated image of (a) reconstructed by the permuted MPS scheme.}
    \label{fig:figure8}
\end{figure}

In this section, we present an example image that demonstrates how our algorithm disentangles a quantum state into a more locally entangled state. 

We use the following convention that interprets a multi-qubit quantum state as binary number:

\begin{equation}
\begin{split}
    \lvert\psi\rangle &= \sum_{q_0,q_2,...q_{n-1}=0}^1 {c_i \lvert q_0 \rangle \otimes \lvert q_1 \rangle  \otimes ... \otimes \lvert q_{n-1} \rangle } \\
    &= \sum_{i=0} ^ {2^n-1} {c_i\lvert i\rangle },
    \end{split}
\end{equation}
where $q_j$ is the state of the j-th qubit and $i=\sum _{j=0} ^ {n-1} 2^{n-1-j} q_j$

Figure \ref{fig:figure8} shows an example permutation for an MPS with the bond dimension of 2, where the local entanglement is maximized. Using our algorithm, we found the optimal permutation to be [3, 4, 2, 7, 9, 6, 8, 5, 0, 1] when 10 qubits are used for amplitude encoding.

As can be observed from Fig. \ref{fig:figure8} (b), there is a clear pattern of spatial repetition that corresponds to more localized entanglement structure. Fig. \ref{fig:figure8} (d) shows the final image reconstructed by reverse permutation after MPS encoding of the data in Fig. \ref{fig:figure8} (b) while Fig. \ref{fig:figure8} (c) represents the image approximated by the standard MPS encoding of the image in Fig. \ref{fig:figure8} (a). Comparison of Fig. \ref{fig:figure8} (c) and (d) demonstrates that the image in Fig. \ref{fig:figure8} (d) retains the essential features and spatial patterns of the original image, indicating that the permutation-based encoding scheme successfully captures the information contained within the image.

\begin{figure}[!t]
    \centering
    \sidesubfloat[]{\includegraphics[width=0.4\textwidth]{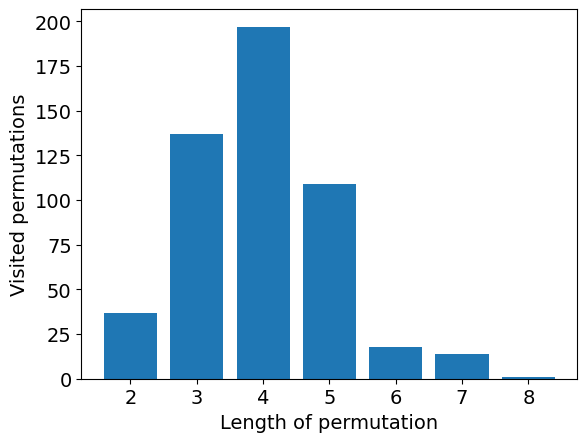}}
    \quad
    \sidesubfloat[]{\includegraphics[width=0.4\textwidth]{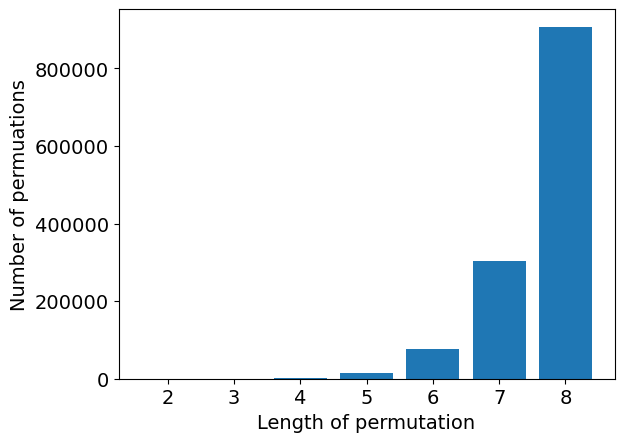}}
    \caption{Comparison of permutation search space. (a) Number of visited nodes during uniform-cost search until finding the optimal permutation corresponding to Fig. \ref{fig:figure8} (b). (b) Size of the entire permutation space.}
    \label{fig:figure9}
\end{figure}

\begin{figure}[!b]
    \centering
    \sidesubfloat[]{\includegraphics[width=0.42\textwidth]{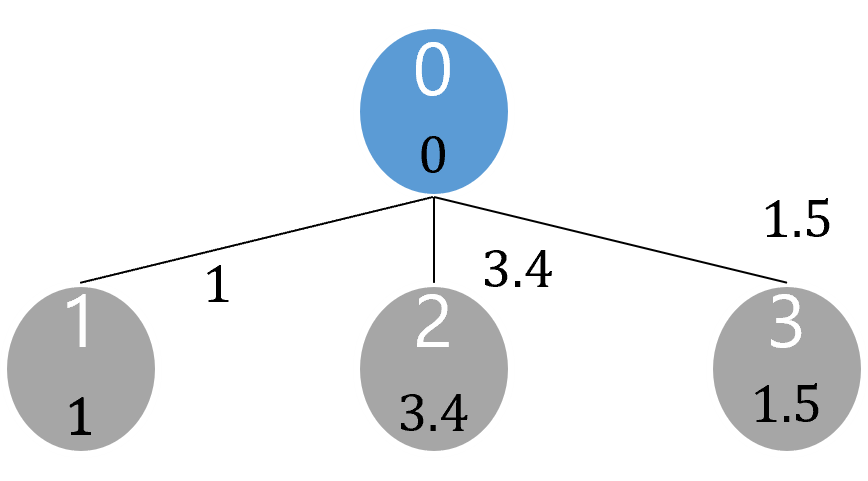}}
    \quad
    \sidesubfloat[]{\includegraphics[width=0.42\textwidth]{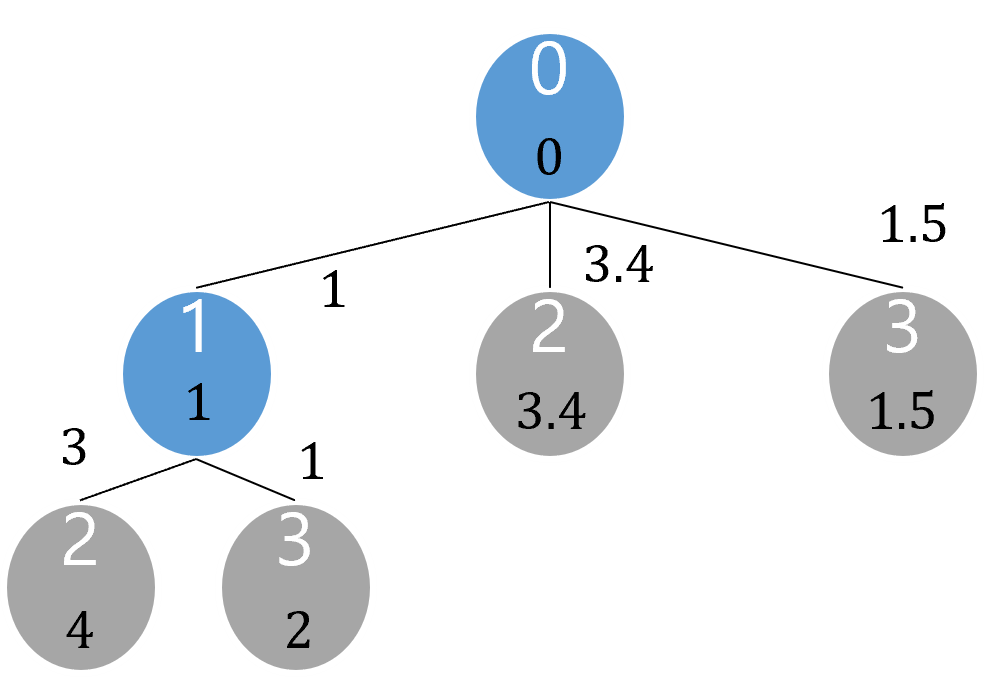}}
    \vskip\baselineskip
    \sidesubfloat[]{\includegraphics[width=0.42\textwidth]{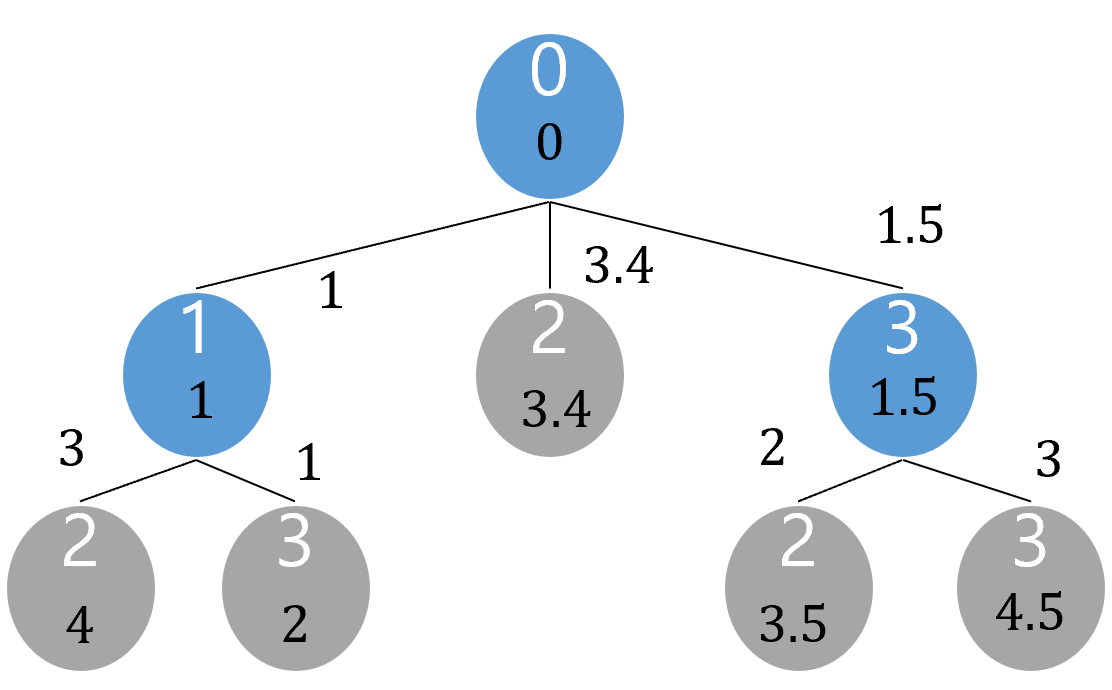}}
    \quad
    \sidesubfloat[]{\includegraphics[width=0.42\textwidth]{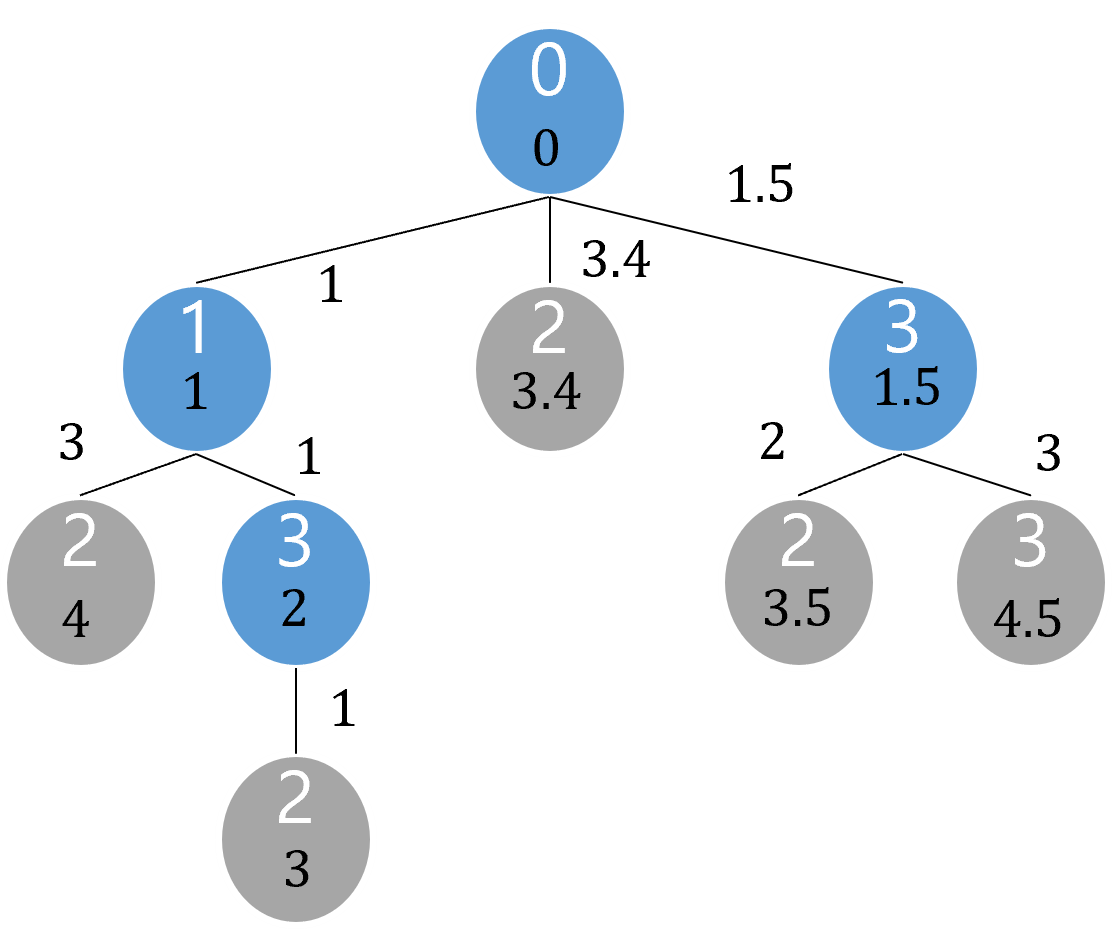}}
    \vskip\baselineskip
    \sidesubfloat[]{\includegraphics[width=0.42\textwidth]{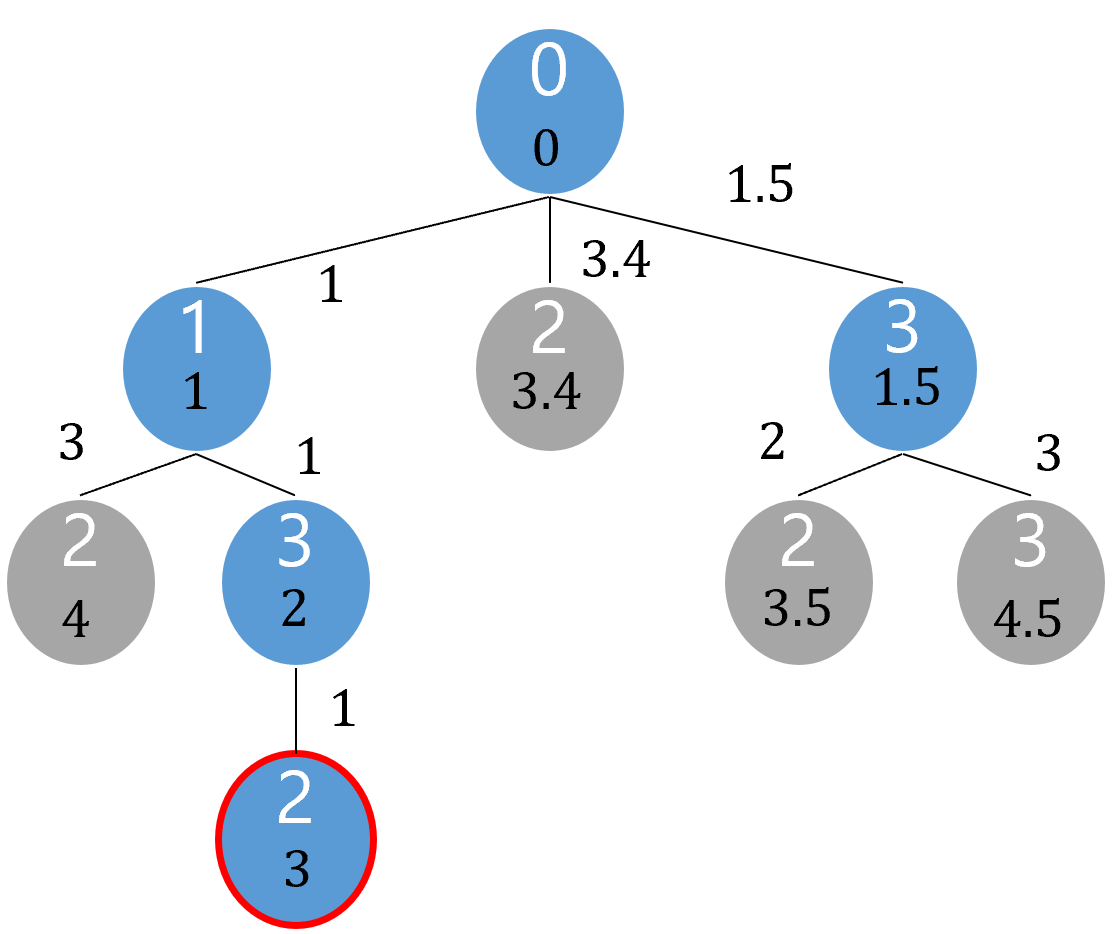}}
    \caption{Illustration of uniform-cost search for the optimal permutation with the lowest truncation error. The visited nodes are represented by blue circles, where the numbers inside each circle indicate the qubit number (white) and the accumulated error (black). The numbers next to the edges represent additional errors associated with the edges. Red circle indicates the optimal permutation. The algorithm always explores the leaf node with the smallest error as follows: (a) pop [0] and push children (b) pop [0,1] and push children (c) pop [0,3] and push children (d) pop [0,1,3] and push children (e) pop [0,1,3,4] and the optimal permutation is found.}
    \label{fig:figure10}
\end{figure}

The search process consists of uniform-cost search of permutations. Since the symmetry mentioned in section 2.2, some permutations yield the same truncation errors, making search space smaller. Therefore, the actual size of search space for length k permutation is not $_{10}{P}_k$, rather $45_{8}{P}_{k-2}$. We plotted visited permutation and actual space size in Fig. \ref{fig:figure9}.

This efficiency of the search process can be attributed to the fact that the uniform-cost search algorithm on a certain type of graph needs to search for permutations that are near optimal. Figure \ref{fig:figure10} illustrates a simplified version of the permutation search with only 4 qubits where we only consider the permutations starting with 0.

For typical images, we noticed that the search process generally terminates quickly after exploring only a small portion of the entire permutation space, demonstrating the practicality of the MPS encoding with permutation.

\section{Image encoding error}
\label{appendixB}

\begin{figure}[!t]
    \centering
    \sidesubfloat[]{\includegraphics[width=0.42\textwidth]{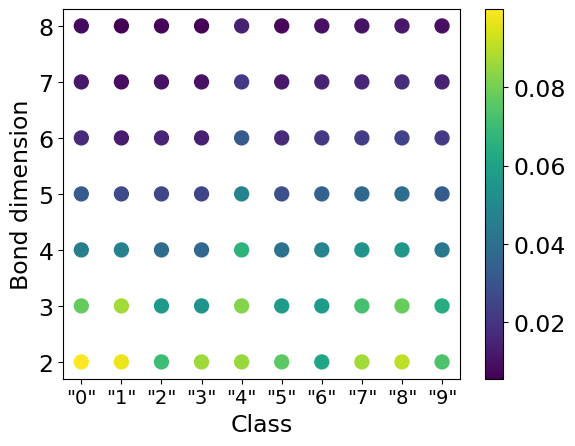}}
    \quad
    \sidesubfloat[]{\includegraphics[width=0.42\textwidth]{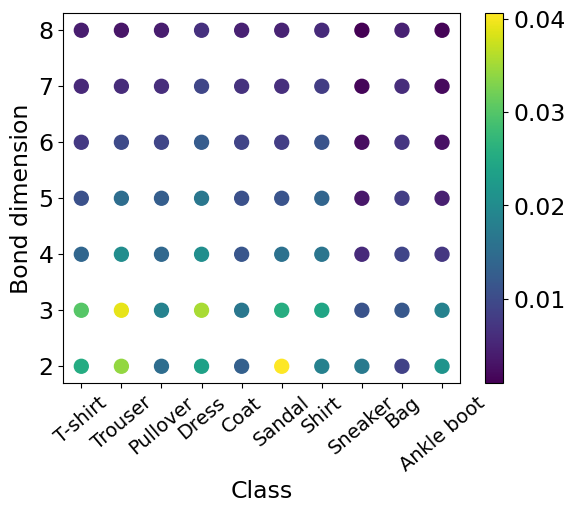}}
    \caption{Difference of Frobenius distances between standard MPS and permuted MPS for each class and bond dimension for (a) MNIST and (b) Fashion-MNIST dataset}
    \label{fig:figure11}
\end{figure}

The encoding error for each class in the MNIST and Fashion-MNIST datasets is plotted in Fig. \ref{fig:figure11} and found that similar patterns emerged: for low bond dimensions, the difference between the Frobenius distance of the standard MPS and that of our algorithm is maximized.

  However, the absolute value of the error was quite different, suggesting that the shape of the image has an impact on the entanglement structure of the amplitude-encoded state, which in turn affects the error.

\section{Quantum network training}
\label{appendixC}

\begin{figure}[!b]
    \centering
    \sidesubfloat[]{\includegraphics[width=0.42\textwidth]{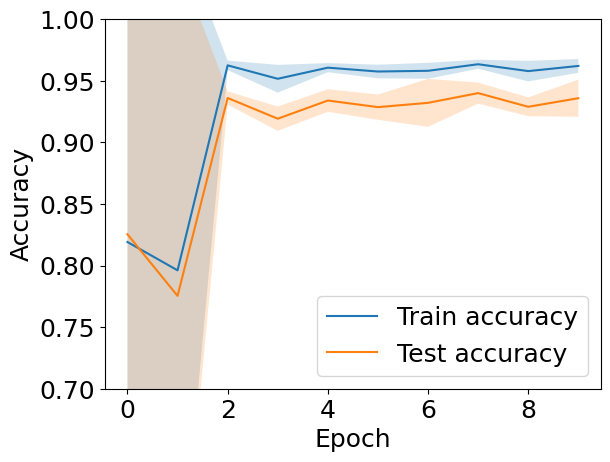}}
    \quad
    \sidesubfloat[]{\includegraphics[width=0.42\textwidth]{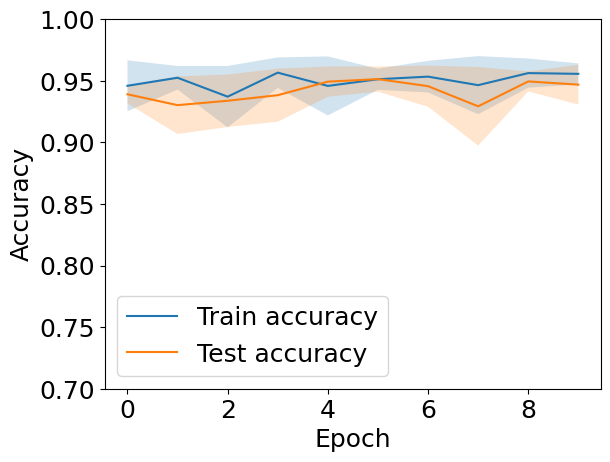}}
    \caption{Training process of the VQC classifiers when the test dataset is encoded by (a) standard MPS and (b) permuted MPS with bond dimension of 2.}
    \label{fig:figure12}
\end{figure}

In Fig. \ref{fig:figure12}, we plot the training process of the VQC classifier used in section 3.1 using two different training datasets: one encoded using standard MPS and the other using our proposed algorithm. Both setups show similar training accuracy, but we observed a clear degradation of test accuracy for the standard MPS, indicating overfitting.

\begin{figure}[!t]
    \centering
    \sidesubfloat[]{\includegraphics[width=0.42\textwidth]{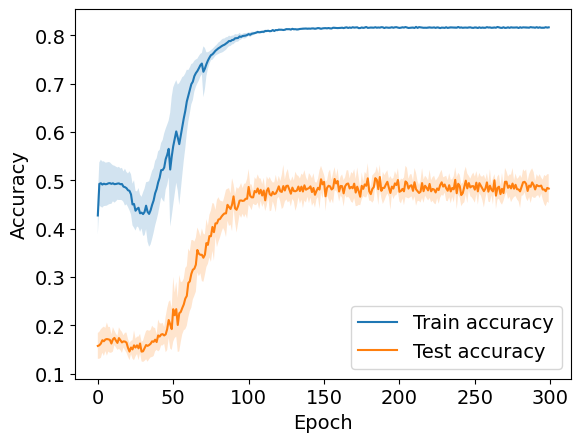}}
    \quad
    \sidesubfloat[]{\includegraphics[width=0.42\textwidth]{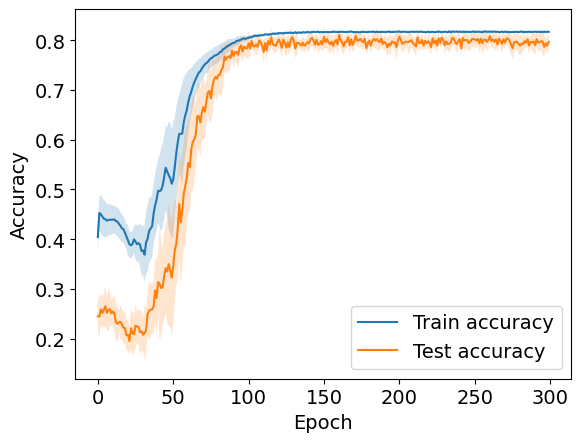}}
    \caption{Training process of the MPS classifiers with (a) standard MPS image and (b) permuted MPS image. The training data is taken from MNIST with bond dimension of 2.}
    \label{fig:figure13}
\end{figure}

Regarding the MPS classifiers shown in section 3.2, Fig. \ref{fig:figure13} also shows a clear separation of test accuracy as observed in the VQC classifier. We noticed that the MPS classifier trained with permuted MPS images converges more quickly.

\section{Discrete cosine transform}
\label{appendixD}

As an alternative to the qubit permutation, we also explored the use of DCT to induce a local entanglement structure in quantum states. In image data, the most significant features are often concentrated in the low-frequency region \cite{RN68}. By applying DCT, we can transform this region into local qubits and therefore expect to observe a higher degree of local entanglement structure.

\begin{figure}[!b]
    \centering
    \sidesubfloat[]{\includegraphics[width=0.19\textwidth]{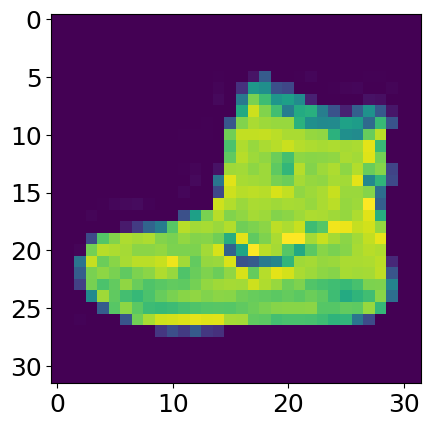}}
    \sidesubfloat[]{\includegraphics[width=0.19\textwidth]{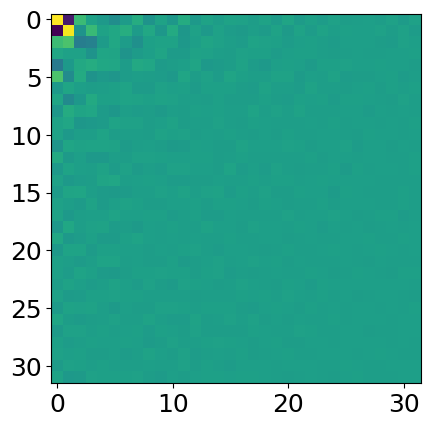}}
    \sidesubfloat[]{\includegraphics[width=0.19\textwidth]{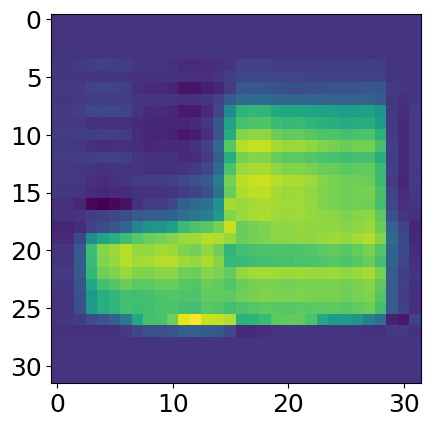}}
    \sidesubfloat[]{\includegraphics[width=0.19\textwidth]{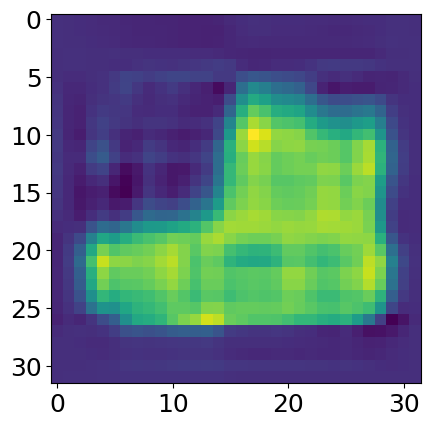}}
    \vskip\baselineskip
    \sidesubfloat[]{\includegraphics[width=0.42\textwidth]{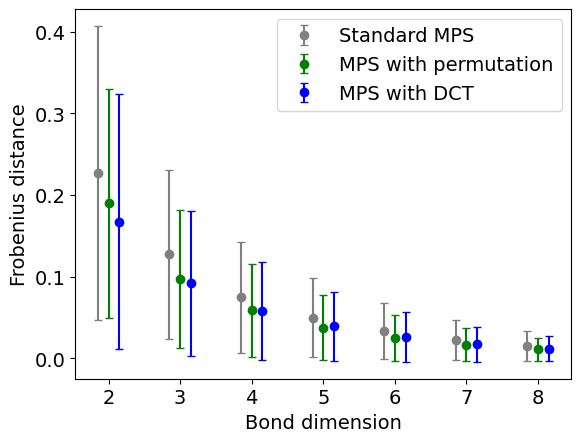}}
    \quad
    \sidesubfloat[]{\includegraphics[width=0.42\textwidth]{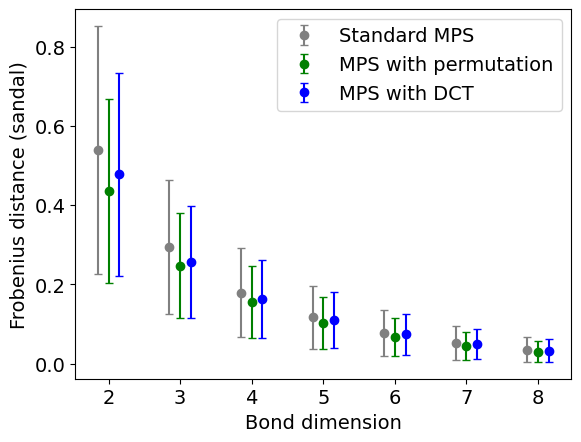}}
    \caption{Comparison of encoding schemes: (a) original image from Fashion-MNIST, (b) frequency components after DCT, (c) standard MPS representation of image with bond dimension of 4, (d) MPS representation with DCT, (e) Frobenius distance comparison of encoding schemes with all the classes in Fashion-MNIST, and (f) Frobenius distance comparison with all the images in the sandal class.}
    \label{fig:figure14}
\end{figure}

Fig. \ref{fig:figure14} (e) shows that DCT does reduce errors similar to the qubit permutation. Note that there exists the dependence of the amount of the reduction in Frobenius distance on the image classes, as illustrated in Fig. \ref{fig:figure14} (f). From a classical computational cost perspective, computing the DCT is generally faster than searching for the optimal qubit permutation. However, inverting the DCT generally requires a quantum circuit \cite{RN69} that can be more complex than inverting a qubit permutation. Therefore, we can conclude that qubit permutation is a more effective approach when preparing quantum states.

\section{Permutation reversal}
\label{appendixE}

\begin{figure}[!t]
    \centering
    \includegraphics[width=0.6\textwidth]{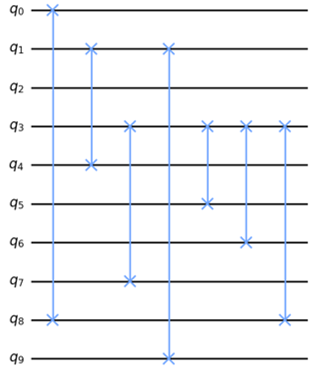}
    \caption{Circuits for permutation reversal using quantum swap gates corresponding to Fig. 8.}
    \label{fig:figure15} 
\end{figure}

Once permutated image state is encoded using the quantum circuit corresponding to the MPS, it is necessary to reverse the permutation in order to obtain the original image state. Typically, this can be achieved simply by qubit renaming. However, in the quantum computer that has limited connectivity, small number of quantum SWAP operations might be necessary. Figure \ref{fig:figure15} illustrates the reversal circuit which incurs negligible overhead in terms of the quantum cost.

\section*{Declaration of Generative AI and AI-assisted technologies in the writing process}
During the preparation of this work the authors used ChatGPT in order to improve readability. After using this tool/service, the authors reviewed and edited the content as needed and take full responsibility for the content of the publication.

\bibliographystyle{elsarticle-num} 
\bibliography{refs}

\begin{thebibliography}{10}
\expandafter\ifx\csname url\endcsname\relax
  \def\url#1{\texttt{#1}}\fi
\expandafter\ifx\csname urlprefix\endcsname\relax\def\urlprefix{URL }\fi
\expandafter\ifx\csname href\endcsname\relax
  \def\href#1#2{#2} \def\path#1{#1}\fi

\bibitem{RN1}
D.~Coppersmith, An approximate fourier transform useful in quantum factoring, IBM Research Report (1994) RC--19642\href {https://doi.org/10.48550/arXiv.quant-ph/0201067} {\path{doi:10.48550/arXiv.quant-ph/0201067}}.

\bibitem{RN2}
P.~W. Shor, Algorithms for quantum computation: discrete logarithms and factoring, in: Proceedings 35th Annual Symposium on Foundations of Computer Science, IEEE, 1994, pp. 124--134.
\newblock \href {https://doi.org/10.1109/SFCS.1994.365700} {\path{doi:10.1109/SFCS.1994.365700}}.

\bibitem{RN3}
A.~W. Harrow, A.~Hassidim, S.~Lloyd, Quantum algorithm for linear systems of equations, Phys. Rev. Lett. 103~(15) (2009) 150502.
\newblock \href {https://doi.org/10.1103/PhysRevLett.103.150502} {\path{doi:10.1103/PhysRevLett.103.150502}}.

\bibitem{duan2020survey}
B.~Duan, J.~Yuan, C.-H. Yu, J.~Huang, C.-Y. Hsieh, A survey on {HHL} algorithm: From theory to application in quantum machine learning, Phys. Lett. A 384~(24) (2020) 126595.
\newblock \href {https://doi.org/https://doi.org/10.1016/j.physleta.2020.126595} {\path{doi:https://doi.org/10.1016/j.physleta.2020.126595}}.

\bibitem{RN4}
P.~Rebentrost, M.~Mohseni, S.~Lloyd, Quantum support vector machine for big data classification, Phys. Rev. Lett. 113~(13) (2014) 130503.
\newblock \href {https://doi.org/10.1103/PhysRevLett.113.130503} {\path{doi:10.1103/PhysRevLett.113.130503}}.

\bibitem{RN5}
S.~Lloyd, M.~Mohseni, P.~Rebentrost, Quantum principal component analysis, Nat. Phys. 10~(9) (2014) 631--633.
\newblock \href {https://doi.org/10.1038/nphys3029} {\path{doi:10.1038/nphys3029}}.

\bibitem{RN6}
I.~Cong, S.~Choi, M.~D. Lukin, Quantum convolutional neural networks, Nat. Phys. 15~(12) (2019) 1273--1278.
\newblock \href {https://doi.org/10.1038/s41567-019-0648-8} {\path{doi:10.1038/s41567-019-0648-8}}.

\bibitem{RN7}
E.~Tang, Quantum principal component analysis only achieves an exponential speedup because of its state preparation assumptions, Phys. Rev. Lett. 127~(6) (2021) 060503.
\newblock \href {https://doi.org/10.1103/PhysRevLett.127.060503} {\path{doi:10.1103/PhysRevLett.127.060503}}.

\bibitem{RN8}
E.~Tang, Dequantizing algorithms to understand quantum advantage in machine learning, Nat. Rev. Phys. 4~(11) (2022) 692--693.
\newblock \href {https://doi.org/10.1038/s42254-022-00511-w} {\path{doi:10.1038/s42254-022-00511-w}}.

\bibitem{RN9}
M.~A. Nielsen, I.~Chuang, Quantum computation and quantum information, American Association of Physics Teachers, 2002.
\newblock \href {https://doi.org/10.1017/CBO9780511976667} {\path{doi:10.1017/CBO9780511976667}}.

\bibitem{RN10}
E.~Knill, R.~Laflamme, G.~J. Milburn, A scheme for efficient quantum computation with linear optics, Nature 409~(6816) (2001) 46--52.
\newblock \href {https://doi.org/10.1038/35051009} {\path{doi:10.1038/35051009}}.

\bibitem{RN11}
I.~F. Araujo, D.~K. Park, F.~Petruccione, A.~J. da~Silva, A divide-and-conquer algorithm for quantum state preparation, Sci. Rep. 11~(1) (2021) 1--12.
\newblock \href {https://doi.org/10.1038/s41598-021-85474-1} {\path{doi:10.1038/s41598-021-85474-1}}.

\bibitem{RN12}
K.~Ghosh, Encoding classical data into a quantum computer (2021).
\newblock \href {http://arxiv.org/abs/2107.09155} {\path{arXiv:2107.09155}}.

\bibitem{RN13}
X.-M. Zhang, M.-H. Yung, X.~Yuan, Low-depth quantum state preparation, Phys. Rev. Res. 3~(4) (2021) 043200.
\newblock \href {https://doi.org/10.1103/PhysRevResearch.3.043200} {\path{doi:10.1103/PhysRevResearch.3.043200}}.

\bibitem{RN14}
S.~Ashhab, Quantum state preparation protocol for encoding classical data into the amplitudes of a quantum information processing register's wave function, Phys. Rev. Res. 4~(1) (2022) 013091.
\newblock \href {https://doi.org/https://doi.org/10.1103/PhysRevResearch.4.013091} {\path{doi:https://doi.org/10.1103/PhysRevResearch.4.013091}}.

\bibitem{RN15}
Z.~Zhao, J.~K. Fitzsimons, P.~Rebentrost, V.~Dunjko, J.~F. Fitzsimons, Smooth input preparation for quantum and quantum-inspired machine learning, Quantum Mach. Intell. 3~(1) (2021) 14.
\newblock \href {https://doi.org/https://doi.org/10.1007/s42484-021-00045-x} {\path{doi:https://doi.org/10.1007/s42484-021-00045-x}}.

\bibitem{RN16}
K.~C. Chen, W.~Dai, C.~Errando-Herranz, S.~Lloyd, D.~Englund, Scalable and high-fidelity quantum random access memory in spin-photon networks, PRX Quantum 2~(3) (2021) 030319.
\newblock \href {https://doi.org/https://doi.org/10.1103/PRXQuantum.2.030319} {\path{doi:https://doi.org/10.1103/PRXQuantum.2.030319}}.

\bibitem{RN17}
V.~Giovannetti, S.~Lloyd, L.~Maccone, Quantum random access memory, Phys. Rev. Lett. 100~(16) (2008) 160501.
\newblock \href {https://doi.org/https://doi.org/10.1103/PhysRevLett.100.160501} {\path{doi:https://doi.org/10.1103/PhysRevLett.100.160501}}.

\bibitem{RN18}
S.~Arunachalam, V.~Gheorghiu, T.~Jochym-O’Connor, M.~Mosca, P.~V. Srinivasan, On the robustness of bucket brigade quantum ram, New J. Phys. 17~(12) (2015) 123010.
\newblock \href {https://doi.org/10.1088/1367-2630/17/12/123010} {\path{doi:10.1088/1367-2630/17/12/123010}}.

\bibitem{RN19}
D.~A. Huffman, A method for the construction of minimum-redundancy codes, Proceedings of the IRE 40~(9) (1952) 1098--1101.
\newblock \href {https://doi.org/https://doi.org/10.1109/JRPROC.1952.273898} {\path{doi:https://doi.org/10.1109/JRPROC.1952.273898}}.

\bibitem{RN20}
T.~A. Welch, A technique for high-performance data compression, Computer 17~(06) (1984) 8--19.
\newblock \href {https://doi.org/https://doi.org/10.1109/MC.1984.1659158} {\path{doi:https://doi.org/10.1109/MC.1984.1659158}}.

\bibitem{RN21}
J.~Rissanen, G.~G. Langdon, Arithmetic coding, IBM J. Res. Dev. 23~(2) (1979) 149--162.
\newblock \href {https://doi.org/https://doi.org/10.1147/rd.232.0149} {\path{doi:https://doi.org/10.1147/rd.232.0149}}.

\bibitem{RN22}
S.~A. Khayam, The discrete cosine transform (dct): theory and application (2003).

\bibitem{RN23}
A.~S. Lewis, G.~Knowles, Image compression using the 2-d wavelet transform, IEEE Trans. Image Process. 1~(2) (1992) 244--250.
\newblock \href {https://doi.org/https://doi.org/10.1109/83.136601} {\path{doi:https://doi.org/10.1109/83.136601}}.

\bibitem{RN24}
R.~Gray, Vector quantization, IEEE ASSP Mag. 1~(2) (1984) 4--29.
\newblock \href {https://doi.org/https://doi.org/10.1109/MASSP.1984.1162229} {\path{doi:https://doi.org/10.1109/MASSP.1984.1162229}}.

\bibitem{RN25}
Y.~Fisher, Fractal image compression: theory and application, Springer Science and Business Media, 2012.
\newblock \href {https://doi.org/https://doi.org/10.1007/978-1-4612-2472-3} {\path{doi:https://doi.org/10.1007/978-1-4612-2472-3}}.

\bibitem{RN26}
I.~Markovsky, Low rank approximation: algorithms, implementation, applications, Vol. 906, Springer, 2012.
\newblock \href {https://doi.org/https://doi.org/10.1007/978-1-4471-2227-2} {\path{doi:https://doi.org/10.1007/978-1-4471-2227-2}}.

\bibitem{RN27}
I.~Markovsky, Structured low-rank approximation and its applications, Automatica 44~(4) (2008) 891--909.
\newblock \href {https://doi.org/https://doi.org/10.1016/j.automatica.2007.09.011} {\path{doi:https://doi.org/10.1016/j.automatica.2007.09.011}}.

\bibitem{RN28}
A.~Novikov, D.~Podoprikhin, A.~Osokin, D.~P. Vetrov, Tensorizing neural networks, in: Advances in Neural Information Processing Systems, Vol.~28, MIT Press, 2015, pp. 442--450.
\newblock \href {https://doi.org/https://doi.org/10.48550/arXiv.1509.06569} {\path{doi:https://doi.org/10.48550/arXiv.1509.06569}}.

\bibitem{RN29}
Z.~Weng, X.~Wang, Low-rank matrix completion for array signal processing, in: 2012 IEEE International Conference on Acoustics, Speech and Signal Processing (ICASSP), IEEE, 2012, pp. 2697--2700.
\newblock \href {https://doi.org/https://doi.org/10.1109/ICASSP.2012.6288473} {\path{doi:https://doi.org/10.1109/ICASSP.2012.6288473}}.

\bibitem{RN30}
D.~Perez-Garcia, F.~Verstraete, M.~M. Wolf, J.~I. Cirac, Matrix product state representations (2006).
\newblock \href {http://arxiv.org/abs/quant-ph/0608197} {\path{arXiv:quant-ph/0608197}}.

\bibitem{RN31}
U.~Schollwöck, The density-matrix renormalization group in the age of matrix product states, Ann. Phys. (N. Y.) 326~(1) (2011) 96--192.
\newblock \href {https://doi.org/https://doi.org/10.1016/j.aop.2010.09.012} {\path{doi:https://doi.org/10.1016/j.aop.2010.09.012}}.

\bibitem{RN32}
R.~Orús, Tensor networks for complex quantum systems, Nat. Rev. Phys. 1~(9) (2019) 538--550.
\newblock \href {https://doi.org/https://doi.org/10.1038/s42254-019-0086-7} {\path{doi:https://doi.org/10.1038/s42254-019-0086-7}}.

\bibitem{RN33}
T.~Ayral, T.~Louvet, Y.~Zhou, C.~Lambert, E.~M. Stoudenmire, X.~Waintal, Density-matrix renormalization group algorithm for simulating quantum circuits with a finite fidelity, PRX Quantum 4~(2) (2023) 020304.
\newblock \href {https://doi.org/https://doi.org/10.1103/PRXQuantum.4.020304} {\path{doi:https://doi.org/10.1103/PRXQuantum.4.020304}}.

\bibitem{RN34}
A.~Muñoz-Moller, L.~Pereira, L.~Zambrano, J.~Cortés-Vega, A.~Delgado, Variational determination of multiqubit geometrical entanglement in noisy intermediate-scale quantum computers, Phys. Rev. Appl. 18~(2) (2022) 024048.
\newblock \href {https://doi.org/https://doi.org/10.1103/PhysRevApplied.18.024048} {\path{doi:https://doi.org/10.1103/PhysRevApplied.18.024048}}.

\bibitem{RN35}
S.~Wouters, D.~Van~Neck, The density matrix renormalization group for ab initio quantum chemistry, Eur. Phys. J. D 68 (2014) 1--20.
\newblock \href {https://doi.org/https://doi.org/10.1140/epjd/e2014-50500-1} {\path{doi:https://doi.org/10.1140/epjd/e2014-50500-1}}.

\bibitem{RN36}
S.-J. Ran, Encoding of matrix product states into quantum circuits of one-and two-qubit gates, Phys. Rev. A 101~(3) (2020) 032310.
\newblock \href {https://doi.org/https://doi.org/10.1103/PhysRevA.101.032310} {\path{doi:https://doi.org/10.1103/PhysRevA.101.032310}}.

\bibitem{RN37}
S.-H. Lin, R.~Dilip, A.~G. Green, A.~Smith, F.~Pollmann, Real-and imaginary-time evolution with compressed quantum circuits, PRX Quantum 2~(1) (2021) 010342.
\newblock \href {https://doi.org/https://doi.org/10.1103/PRXQuantum.2.010342} {\path{doi:https://doi.org/10.1103/PRXQuantum.2.010342}}.

\bibitem{RN38}
P.~Gundlapalli, J.~Lee, Deterministic and entanglement-efficient preparation of amplitude-encoded quantum registers, Phys. Rev. Appl. 18~(2) (2022) 024013.
\newblock \href {https://doi.org/https://doi.org/10.1103/PhysRevApplied.18.024013} {\path{doi:https://doi.org/10.1103/PhysRevApplied.18.024013}}.

\bibitem{RN39}
L.~Deng, The mnist database of handwritten digit images for machine learning research [best of the web], IEEE Signal Process. Mag. 29~(6) (2012) 141--142.
\newblock \href {https://doi.org/https://doi.org/10.1109/MSP.2012.2211477} {\path{doi:https://doi.org/10.1109/MSP.2012.2211477}}.

\bibitem{RN40}
H.~Xiao, K.~Rasul, R.~Vollgraf, Fashion-mnist: a novel image dataset for benchmarking machine learning algorithms (2017).
\newblock \href {http://arxiv.org/abs/1708.07747} {\path{arXiv:1708.07747}}.

\bibitem{RN41}
R.~Dilip, Y.-J. Liu, A.~Smith, F.~Pollmann, \href{<Go to ISI>://WOS:000876478200007}{Data compression for quantum machine learning}, Phys. Rev. Res. 4~(4) (2022) 043007.
\newblock \href {https://doi.org/https://doi.org/10.1103/PhysRevResearch.4.043007} {\path{doi:https://doi.org/10.1103/PhysRevResearch.4.043007}}.
\newline\urlprefix\url{<Go to ISI>://WOS:000876478200007}

\bibitem{RN42}
L.~Wright, F.~Barratt, J.~Dborin, V.~Wimalaweera, B.~Coyle, A.~Green, Deterministic tensor network classifiers (2022).
\newblock \href {http://arxiv.org/abs/2205.09768} {\path{arXiv:2205.09768}}.

\bibitem{RN43}
S.~Efthymiou, J.~Hidary, S.~Leichenauer, Tensornetwork for machine learning, arXiv:1906.06329 [cs.LG] (2019).
\newblock \href {http://arxiv.org/abs/1906.06329} {\path{arXiv:1906.06329}}.

\bibitem{RN44}
J.~Liu, S.~Li, J.~Zhang, P.~Zhang, Tensor networks for unsupervised machine learning, Phys. Rev. E 107~(1) (2023) L012103.
\newblock \href {https://doi.org/https://doi.org/10.1103/PhysRevE.107.L012103} {\path{doi:https://doi.org/10.1103/PhysRevE.107.L012103}}.

\bibitem{RN45}
A.~Krizhevsky, Learning multiple layers of features from tiny images, Master, Univ. of Toronto (2009).

\bibitem{RN46}
I.~V. Oseledets, Tensor-train decomposition, SIAM J. Sci. Comput. 33~(5) (2011) 2295--2317.
\newblock \href {https://doi.org/https://doi.org/10.1137/090752286} {\path{doi:https://doi.org/10.1137/090752286}}.

\bibitem{RN47}
R.~Orús, A practical introduction to tensor networks: Matrix product states and projected entangled pair states, Ann. Phys. (N. Y.) 349 (2014) 117--158.
\newblock \href {https://doi.org/https://doi.org/10.1016/j.aop.2014.06.013} {\path{doi:https://doi.org/10.1016/j.aop.2014.06.013}}.

\bibitem{RN48}
S.~Wold, K.~Esbensen, P.~Geladi, Principal component analysis, Chemom. Intell. Lab. Syst. 2~(1-3) (1987) 37--52.
\newblock \href {https://doi.org/https://doi.org/10.1016/0169-7439(87)80084-9} {\path{doi:https://doi.org/10.1016/0169-7439(87)80084-9}}.

\bibitem{RN49}
S.~J. Russell, Artificial intelligence a modern approach, Pearson Education, Inc., 2010.

\bibitem{RN50}
E.~Dijkstra, A note on two problems in connexion with graphs, Numer. Math. 1 (1959) 269--271.
\newblock \href {https://doi.org/https://doi.org/10.1007/BF01386390} {\path{doi:https://doi.org/10.1007/BF01386390}}.

\bibitem{RN51}
H.~Jeon, Mps search, \url{https://github.com/snu-quiqcl/MPS-search}.

\bibitem{RN52}
R.~LaRose, B.~Coyle, Robust data encodings for quantum classifiers, Phys. Rev. A 102~(3) (2020) 032420.
\newblock \href {https://doi.org/https://doi.org/10.1103/PhysRevA.102.032420} {\path{doi:https://doi.org/10.1103/PhysRevA.102.032420}}.

\bibitem{RN53}
S.~Lloyd, M.~Schuld, A.~Ijaz, J.~Izaac, N.~Killoran, Quantum embeddings for machine learning, arXiv:2001.03622 [quant-ph] (2020).
\newblock \href {http://arxiv.org/abs/2001.03622} {\path{arXiv:2001.03622}}.

\bibitem{li2022quantum}
L.~H. Li, D.-B. Zhang, Z.~Wang, Quantum kernels with gaussian state encoding for machine learning, Phys. Lett. A 436 (2022) 128088.
\newblock \href {https://doi.org/https://doi.org/10.1016/j.physleta.2022.128088} {\path{doi:https://doi.org/10.1016/j.physleta.2022.128088}}.

\bibitem{RN54}
X.~Yao, H.~Wang, Z.~Liao, M.-C. Chen, J.~Pan, J.~Li, K.~Zhang, X.~Lin, Z.~Wang, Z.~Luo, Quantum image processing and its application to edge detection: theory and experiment, Phys. Rev. X 7~(3) (2017) 031041.
\newblock \href {https://doi.org/https://doi.org/10.1103/PhysRevX.7.031041} {\path{doi:https://doi.org/10.1103/PhysRevX.7.031041}}.

\bibitem{RN55}
Q.~Bai, X.~Hu, Quantity study on a novel quantum neural network with alternately controlled gates for binary image classification, Quantum Inf. Process. 22~(5) (2023) 184.
\newblock \href {https://doi.org/https://doi.org/10.1007/s11128-023-03929-y} {\path{doi:https://doi.org/10.1007/s11128-023-03929-y}}.

\bibitem{RN56}
J.~Biamonte, P.~Wittek, N.~Pancotti, P.~Rebentrost, N.~Wiebe, S.~Lloyd, Quantum machine learning, Nature 549~(7671) (2017) 195--202.
\newblock \href {https://doi.org/https://doi.org/10.1038/nature23474} {\path{doi:https://doi.org/10.1038/nature23474}}.

\bibitem{RN57}
M.~S. Rudolph, J.~Chen, J.~Miller, A.~Acharya, A.~Perdomo-Ortiz, Decomposition of matrix product states into shallow quantum circuits, arXiv:2209.00595 [quant-ph] (2022).
\newblock \href {http://arxiv.org/abs/2209.00595} {\path{arXiv:2209.00595}}.

\bibitem{RN58}
A.~Mari, T.~R. Bromley, J.~Izaac, M.~Schuld, N.~Killoran, Transfer learning in hybrid classical-quantum neural networks, Quantum 4 (2020) 340.
\newblock \href {https://doi.org/https://doi.org/10.22331/q-2020-10-09-340} {\path{doi:https://doi.org/10.22331/q-2020-10-09-340}}.

\bibitem{RN59}
E.~Ovalle-Magallanes, J.~G. Avina-Cervantes, I.~Cruz-Aceves, J.~Ruiz-Pinales, Hybrid classical–quantum convolutional neural network for stenosis detection in x-ray coronary angiography, Expert Syst. Appl. 189 (2022) 116112.
\newblock \href {https://doi.org/https://doi.org/10.1016/j.eswa.2021.116112} {\path{doi:https://doi.org/10.1016/j.eswa.2021.116112}}.

\bibitem{RN60}
K.~He, X.~Zhang, S.~Ren, J.~Sun, Deep residual learning for image recognition, in: Proceedings of the IEEE Conference on Computer Vision and Pattern Recognition, 2015.
\newblock \href {http://arxiv.org/abs/1512.03385} {\path{arXiv:1512.03385}}, \href {https://doi.org/https://doi.org/10.48550/arXiv.1512.03385} {\path{doi:https://doi.org/10.48550/arXiv.1512.03385}}.

\bibitem{RN61}
chenyaofo, Pytorch cifar models, GitHub repository (2021).

\bibitem{RN62}
D.~P. Kingma, J.~Ba, Adam: A method for stochastic optimization, arXiv:1412.6980 [cs.LG] (2014).
\newblock \href {http://arxiv.org/abs/1412.6980} {\path{arXiv:1412.6980}}.

\bibitem{RN63}
G.~Hellstem, Hybrid quantum network for classification of finance and mnist data, in: 2021 IEEE 18th International Conference on Software Architecture Companion (ICSA-C), IEEE, 2021, pp. 1--4.
\newblock \href {https://doi.org/10.1109/ICSA-C52384.2021.00027} {\path{doi:10.1109/ICSA-C52384.2021.00027}}.

\bibitem{RN64}
I.~J. Good, Rational decisions, J. Roy. Statist. Soc. Ser. B 14~(1) (1952) 107--114.
\newblock \href {https://doi.org/https://doi.org/10.1111/j.2517-6161.1952.tb00104.x} {\path{doi:https://doi.org/10.1111/j.2517-6161.1952.tb00104.x}}.

\bibitem{RN65}
V.~Bergholm, J.~Izaac, M.~Schuld, C.~Gogolin, S.~Ahmed, V.~Ajith, M.~S. Alam, G.~Alonso-Linaje, B.~AkashNarayanan, A.~Asadi, Pennylane: Automatic differentiation of hybrid quantum-classical computations, arXiv:1811.04968 [quant-ph] (2018).
\newblock \href {http://arxiv.org/abs/1811.04968} {\path{arXiv:1811.04968}}.

\bibitem{RN66}
L.~Banchi, J.~Pereira, S.~Pirandola, Generalization in quantum machine learning: A quantum information standpoint, PRX Quantum 2~(4) (2021) 040321.
\newblock \href {https://doi.org/https://doi.org/10.1103/PRXQuantum.2.040321} {\path{doi:https://doi.org/10.1103/PRXQuantum.2.040321}}.

\bibitem{RN67}
J.~Preskill, Quantum computing in the nisq era and beyond, Quantum 2 (2018) 79.
\newblock \href {https://doi.org/https://doi.org/10.22331/q-2018-08-06-79} {\path{doi:https://doi.org/10.22331/q-2018-08-06-79}}.

\bibitem{RN68}
G.~K. Wallace, The jpeg still picture compression standard, IEEE Trans. Consum. 38~(1) (1992) xviii--xxxiv.
\newblock \href {https://doi.org/10.1109/30.125072} {\path{doi:10.1109/30.125072}}.

\bibitem{RN69}
A.~Klappenecker, M.~Rotteler, Discrete cosine transforms on quantum computers, in: Proceedings of the 2nd International Symposium on Image and Signal Processing and Analysis, IEEE, 2001, pp. 464--468.
\newblock \href {https://doi.org/https://doi.org/10.1109/ISPA.2001.938674} {\path{doi:https://doi.org/10.1109/ISPA.2001.938674}}.

\end{thebibliography}

\end{document}